\documentclass[fleqn,usenatbib]{mnras}

\usepackage{caption} 
\usepackage{newtxtext,newtxmath}

\usepackage[T1]{fontenc}

\DeclareRobustCommand{\VAN}[3]{#2}
\let\VANthebibliography\thebibliography
\def\thebibliography{\DeclareRobustCommand{\VAN}[3]{##3}\VANthebibliography}

\usepackage{graphicx}	
\usepackage{amsmath}

\title[Shao et al.]{The Spectral Behaviour and Variability of Narrow-line Seyfert 1 Galaxies with Australia Telescope Compact Array Observations}

\author[]{
Xi Shao,$^{1,2,3}$
Philip G. Edwards,$^{3}$\thanks{E-mail: Philip.Edwards@csiro.au}
Jamie Stevens,$^{4}$
Minfeng Gu,$^{1}$
Timothy J. Galvin,$^{5}$
and Minh T. Huynh$^{5}$ 
\\
$^{1}$ Shanghai Astronomical Observatory, Chinese Academy of Sciences, 80 Nandan Road, Shanghai 200030, People’s Republic of China\\
$^{2}$ University of Chinese Academy of Sciences, 19A Yuquan Road, Beijing 100049, People’s Republic of China\\
$^{3}$ CSIRO Space and Astronomy, Australian Telescope National Facility, Epping, New South Wales 1710, Australia\\
$^{4}$ CSIRO Space and Astronomy, 1828 Yarrie Lake Road, Narrabri, NSW 2390, Australia\\
$^{5}$ CSIRO Space and Astronomy, PO Box 1130, Bentley, WA 6102, Australia}

\date{Accepted XXX. Received YYY; in original form ZZZ}

\pubyear{2024}

\begin{document}
\label{firstpage}
\pagerange{\pageref{firstpage}--\pageref{lastpage}}
\maketitle

\begin{abstract}
We present multi-frequency radio data for a sample of narrow-line Seyfert~1 galaxies. 
We first focus on the sub-class of gamma-ray emitting narrow-line Seyfert~1 galaxies, studying the long-term radio variability of five sources and comparing it to their gamma-ray state.   
We then extend the observations of the southern narrow-line Seyfert~1 galaxy sample of Chen et al.\ 
by observing several candidate narrow-line Seyfert~1 sources for the first time, and re-observing several other gamma-ray quiet sources to obtain a first indication of their radio variability. 
We find that the gamma-ray emitting narrow-line Seyfert~1 galaxies are highly variable radio emitters and that there
are instances of contemporaneous flaring activity between the radio and gamma-ray band (PKS 0440$-$00, PMN J0948+0022 and PKS 1244$-$255). However, there are also cases of significant radio outbursts without gamma-ray counterparts (PMN J0948+0022 and PKS 2004$-$447). The five gamma-ray NLS1s favour flat or inverted radio spectra, although the spectral indices vary significantly over time. For the gamma-ray quiet sample, the difference between the previous observations at 5.5\,GHz and new ATCA observations indicates that over half of the 14 sources exhibit apparent variability. In contrast to gamma-ray loud sources, gamma-ray quiet objects tend to have steep spectra especially in the lower radio band (887.5--1367.5\,MHz), with a number of the variable sources having flatter spectra at higher radio frequencies. 
\end{abstract}




\section{Introduction} \label{sec:intro}

Seyfert galaxies are a class of active galactic nuclei (AGN) that were first identified based on their appearance as otherwise normal galaxies with an abnormally bright central core.
Spectroscopically, Seyferts can be subdivided into several sub-types.
Type~1 Seyfert galaxies
have the widths of permitted emission lines 
Doppler broadened to $\sim$5000 km\,s$^{-1}$ but with
forbidden lines much narrower, with widths typically $\sim$500 km\,s$^{-1}$.
A further sub-class is the 
narrow-line Seyfert~1 galaxies (NLS1s) that were first described by \citet{1985ApJ...297..166O}. 
These have, as their name suggests, somewhat narrower lines from the broad line region ($\rm{FWHM(H\beta)} <2000\,\rm{km\,s^{-1}}$) and have a ratio of [O III] to $\rm H{\beta}$ emission lines of less than 3. 
They also typically have relatively strong Fe II emission \citep{1989ApJ...342..224G}. These characteristics distinguish NLS1s from broad-line Seyfert 1 (BLS1) galaxies. 

In contrast to BLS1s, it is generally the case that NLS1s have smaller supermassive black hole masses, in the range $\sim 10^6-10^8 M_{\odot}$ \citep{1999ApJ...521L..95P,2011arXiv1109.4181P,2015A&A...573A..76J,2016MNRAS.462.1256C}. (However, \citet{2019ApJ...881L..24V} has found that NLS1s have black hole masses and accretion rates similar to BLS1s.)
NLS1s have high Eddington ratios, ranging from 0.1 to 1, and even larger than 1 in some cases \citep[e.g.][]{1992ApJS...80..109B,2018FrASS...5....6M,2022MNRAS.509.3599T}.
  
The majority of NLS1s are hosted in spiral galaxies, but some are  found in disk-like galaxies \citep[e.g.][]{2018A&A...619A..69J,2020MNRAS.492.1450O,2022A&A...668A..91V,2023A&A...679A..32V} or having elliptical hosts \citep{2017MNRAS.469L..11D,2018MNRAS.478L..66D}.

The most extensive catalogue of candidate NLS1 galaxies is that compiled by \cite{Paliya2024}, based on a decomposition of optical spectra from the Sloan Digital Sky Survey Data Release~17. This catalogue contains 22656 NLS1 galaxies, with a companion catalogue of 52273 broad-line Seyfert 1 (BLS1) galaxies. 

At X-ray energies, NLS1s generally show strong flux variability \citep{1999ApJ...526...52T} and steeper spectral indices \citep{1998A&A...330...25G,1999ApJS..125..317L} with respect to BLS1s.
Many NLS1s have a prominent soft X-ray excess \citep{1996A&A...305...53B,1999ApJS..125..317L,2006MNRAS.365.1067C,2006ApJS..166..128Z},
with this characteristic proving to be an effective way of identifying new NLS1s.
Possible origins include thermal emission from the accretion disc \citep{1989MNRAS.240..833T}, non-thermal emission from the jet \citep{2000ApJ...533..650S}, or the Comptonization of photons from disc or broad-line region (BLR) \citep{2007MNRAS.375..417C}.

The {\em Compton Gamma Ray Observatory} did not detect gamma-rays from any known NLS1s, and so it was something of a surprise when {\em Fermi} detected gamma-rays from PMN J0948+0022 in the first year of the mission \citep{Abdo2009}.
The variability at gamma-ray energies of PMN J0948+0022, and subsequently detected NLS1s, excluded a possible starburst origin for the gamma-ray photons and confirmed the presence of relativistic jets in these sources \citep{Calderone2011}.
Recently, \cite{Foschini2022} have systematically compiled
a new sample of gamma-ray emitting jetted AGN
from fourth catalogue of gamma-ray point sources produced
by the {\it Fermi} Large Area Telescope (LAT) \citep{2020ApJS..247...33A}.
They examined available
optical spectra of 2980 gamma-ray point sources 
to measure redshifts and to confirm or change the original
LAT classification. 
This reclassification resulted in 24 NLS1 confirmed sources or candidates.
\cite{Foschini2022} note that although NLS1s lie at the 
low-luminosity tail of the flat-spectrum radio quasar (FSRQ) distribution, the 
generally smaller supermassive black hole mass and 
high accretion rate are key differences, making
this sub-class of particular interest.
Several sources listed as NLS1s in the Fourth {\it Fermi}-LAT source catalogue (4FGL) were reclassified by \cite{Foschini2022}, and conversely, a number of sources listed as flat-spectrum radio quasars in 4FGL were reclassified by \cite{Foschini2022} as NLS1s.

In Table\,\ref{tab:fermi_NLS1s} we list all sources included in the 4FGL catalogue that have been associated with a NLS1.
These are considered in detail by \cite{Foschini2022}, with the fifth column of the table listing their classification of the source. The final grouping of candidates have been suggested as candidate NLS1s by others \citep{2015MNRAS.454L..16Y,2017FrASS...4....8B,2018ApJ...853L...2P,2019MNRAS.487L..40Y}.
(We note that additionally, several sources not listed in 4FGL, and therefore not included in Table~1, have been proposed as gamma-ray NLS1s: including
SDSS J0031+0936 \citep{2019ApJ...871..162P},
FBQS J1102+2239 \citep{2011nlsg.confE..24F},
TXS 1518+423 \citep{2018ApJ...853L...2P}, 
and 
SDSS J1641+3454 \citep{2018A&A...614L...1L}.)

Consistent with the inference that variable gamma-ray emitting NLS1 galaxies have jets, the majority of this sub-class display a core-jet morphology on the parsec-scale or kiloparsec-scale \citep[e.g.,][]
{2018Lister,2019Galax...7...87D,2023ApJ...943..136S}.
This branch of NLS1s provide a good chance to study jet evolution at an early stage \citep[e.g.,][]{2023MNRAS.523..441Y}.

It is traditional to consider AGN based on their radio loudness, $R_L$, 
defined by the flux density ratio of rest-frame 
5\,GHz and 4400~\AA  \citep{1989AJ.....98.1195K}
Most NLS1s are radio quiet (RQ), with $R_L < 10$, or radio silent (RS),
with only 5$\sim$7\% classified as radio loud (RL). 
However, it has been noted that the usefulness of radio loudness in classifying NLS1s may be limited \citep[e.g.][]{2011RAA....11.1266F,2017A&A...606A...9J,2018Lister,2020A&A...636A..64B}.
However, in general terms,
radio-loud NLS1 sources have flat spectra in the 1.4--5\,GHz range, with some displaying inverted spectra
\citep[e.g.,][]{2015ApJS..221....3G}.
In contrast, most radio-quiet NLS1 sources have steep
spectra, either as a power-law, or curved spectra becoming increasingly steep with increasing frequency \citep[e.g.][]{2022MNRAS.512..471C}.

\cite{2006AJ....132..531K}
found that 35 of the 128 (27\%) NLS1s they considered
had catalogued radio counterparts, but as noted by \cite{2018Lister}, 
all-sky catalogues are generally not very deep.
This was demonstrated by the targeted radio observations of 
southern NLS1 candidates, which resulted in the detection of
49 of 62 sources (79\%) with the Karl G. Jansky Very Large Array (VLA) at 
5.5\,GHz \citep{2020MNRAS.498.1278C} 
and 42 of 85 (49\%) with the Australia Telescope Compact Array (ATCA) at 5.5 and/or 9.0\,GHz \citep{2022MNRAS.512..471C}, respectively.

In this paper, we explore the spectral behaviour and variability of NLS1s with archival ATCA radio data and new observations. 
In Section~\ref{sec:gamma-loud} we focus on the sub-class of gamma-ray emitting NLS1s, studying the long-term radio variability of five sources and comparing it to their gamma-ray state.   
In Section~\ref{sec:gamma-quiet} we extend the observations of \citet{2018A&A...615A.167C} by observing several candidate NLS1 sources for the first time, and re-observing several other gamma-ray quiet sources to obtain a first indication of their radio variability. 
We discuss the difference between our observing results and the previous observations and the underlying emission mechanism in Section~\ref{discussion}.
The work is summarised in Section~\ref{conclusion}. 
Throughout the paper, we follow the convention $S_\nu\propto \nu^{\alpha}$ where $S$ is flux density, $\nu$ is frequency, and $\alpha$ is the spectral index.

\section{Gamma-ray loud NLS1 galaxies} \label{sec:gamma-loud}

There is long-term ATCA monitoring for five 
$Fermi$ $\gamma$-ray sources in Table~\ref{tab:fermi_NLS1s}:
two established NLS1s (PMN J0948+0022 and PKS 1502+036),
two sources classified as NLS1 by \cite{Foschini2022} 
(PKS 0440$-$00 and PKS 1244$-$255) and the
candidate NLS1, classified as a misaligned AGN by
\cite{Foschini2022}, PKS 2004$-$447.

We compare ATCA multi-frequency light curves (using data primarily from ATCA projects C007 and C1730) with {\it Fermi} light curves.
The ATCA project C007 is an Observatory-led program to monitor several hundred bright AGN at multiple frequencies several times per year so that observers can assess their usefulness as calibrator sources. The C1730 project specifically monitors a smaller number of selected gamma-ray sources at a higher cadence, again at multiple frequencies. Data from both programs is pipeline-processed with the results being made available through the ATCA Calibrator Database\footnote{https://www.narrabri.atnf.csiro.au/calibrators/}.
Note that the reported errors here are statistical only: systematic errors of up to 5\% may also be present at 5.5 and 9\, GHz \citep[see, e.g.,][]{Tingay2003}, and up to 10\% at higher frequencies
(where atmospheric opacity effects and antenna pointing become more important). 

The {\it Fermi} data used here is taken from the {\it Fermi}-LAT Light Curve Repository (LCR)
\citep{2023abdollahi}.
The LCR is a database of flux-calibrated light curves for over 1500 sources deemed to be variable in the 10 year {\it Fermi}-LAT point source (4FGL-DR2) catalogue \citep{2020arXiv200511208B}. We have used data binned in 1~week intervals in this paper.

\subsubsection{PKS\,0440$-$00} \label{sec:0440-00}

\begin{figure*}
	\includegraphics[width=2\columnwidth]{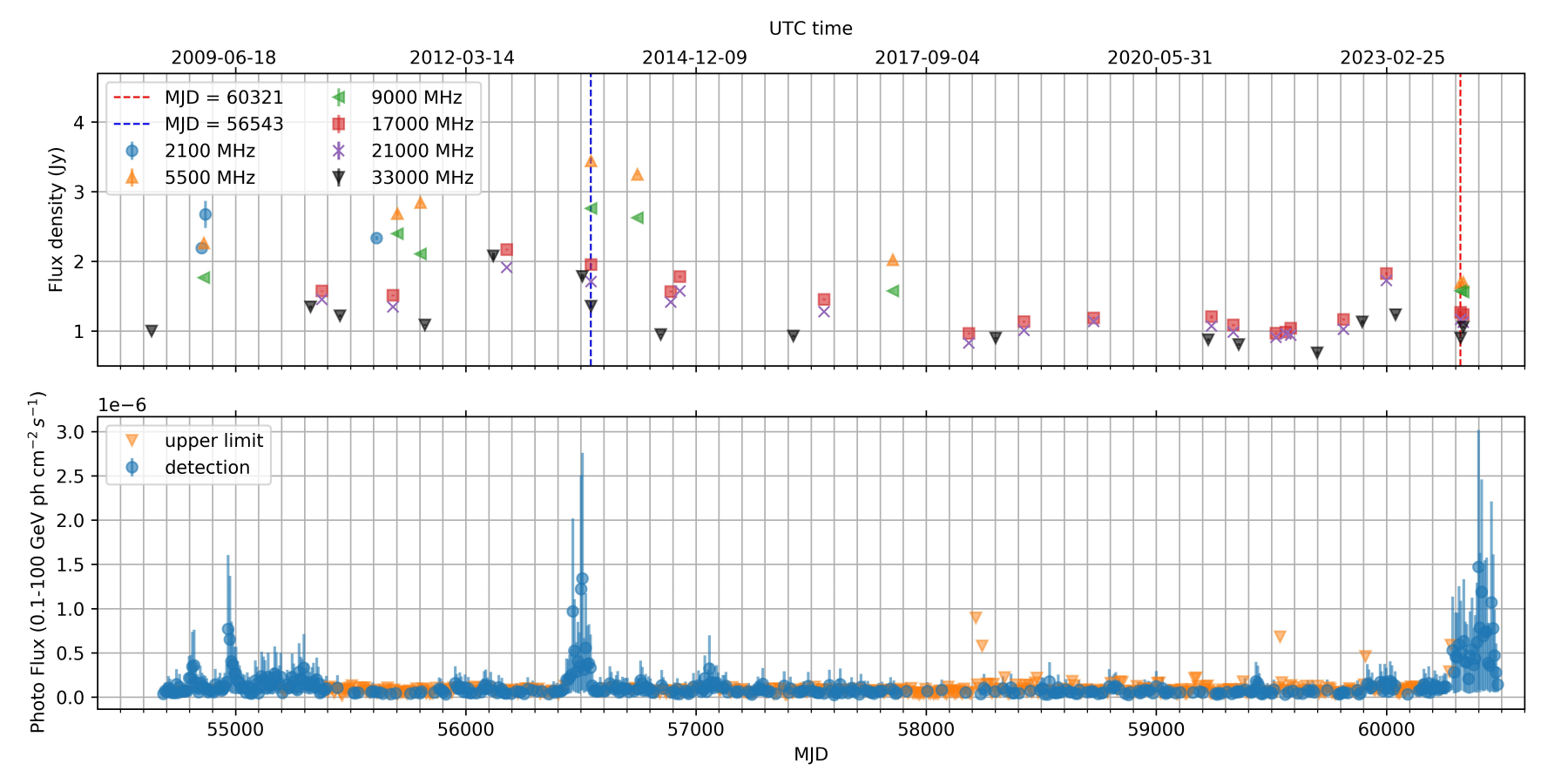}
    \caption{The radio and gamma-ray light curves of PKS 0440$-$00. The dashed lines indicate the epochs for which radio spectra are shown in Figure~\ref{fig:spectra}.}
    \label{fig:0440-003}
\end{figure*}

The {\it Fermi} light curve for PKS\,0440$-$00 in Fig.~\ref{fig:0440-003} shows three epochs of enhanced activity,
centred on MJD 55000, 56500 and 60200.
The radio monitoring is sparse, but it is notable that
the highest 5.5 and 9.0 GHz flux densities approximately
coincide with the MJD 56500 gamma-ray outburst, and the
highest 17, 21 and 33\,GHz flux densities precede it.
The last epoch of radio monitoring occurred shortly after the
MJD 60200 gamma-ray outburst had started: the radio
flux densities were not particularly high, but the spectrum
had flattened (see \S 2.2).

\subsubsection{PMN\,J0948+0022} \label{sec:J0948+0022}

\begin{figure*}
        \includegraphics[width=2\columnwidth]{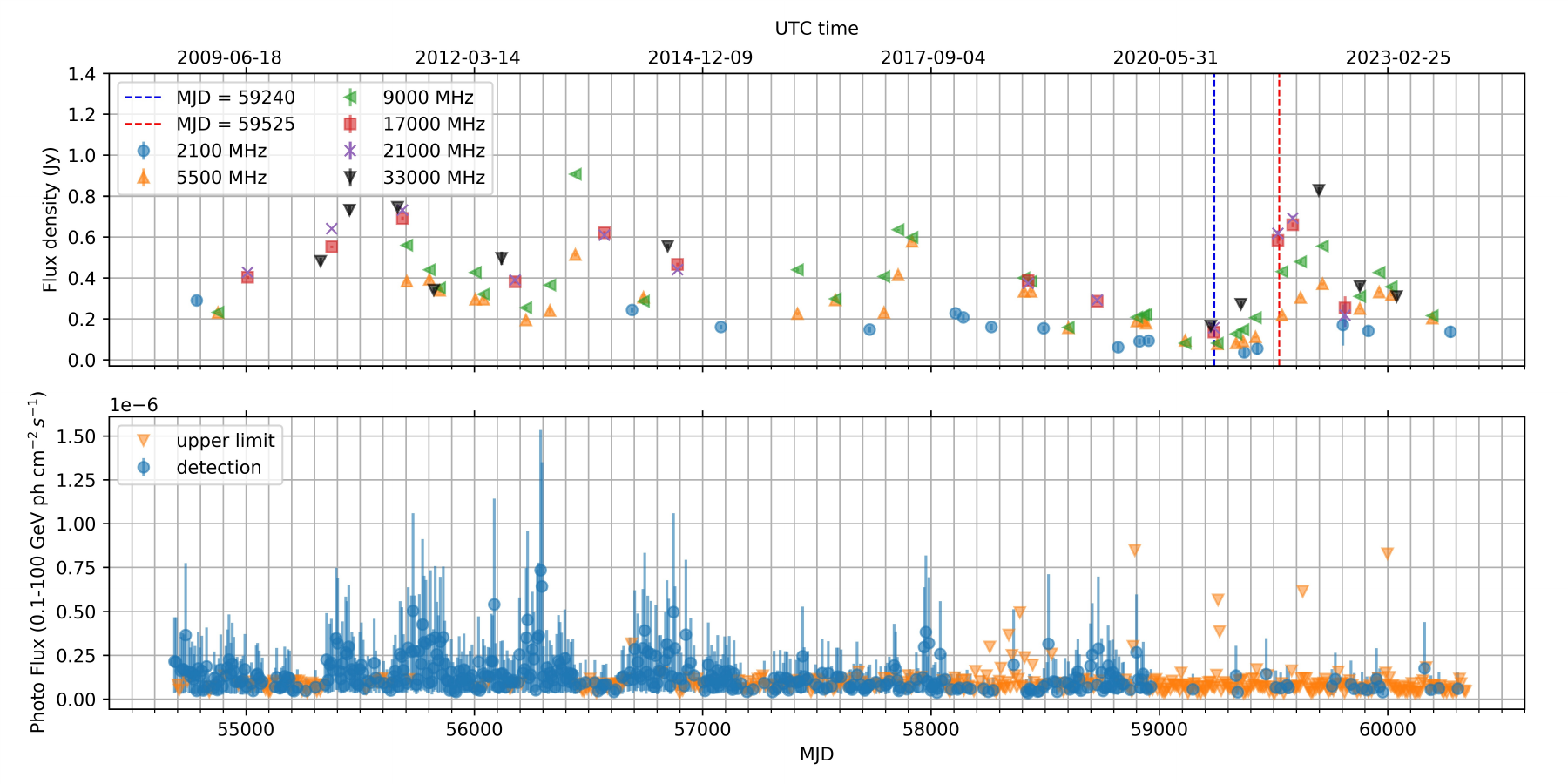}
    \caption{The radio and gamma-ray light curves of PMN J0948+0022. The dashed lines indicate the epochs for which radio spectra are shown in Figure~\ref{fig:spectra}.}
    \label{fig:0946+006}
\end{figure*}

PMN\,J0948+0022 was the first NLS1 detected by {\it Fermi}, and the light curve
in Fig.~\ref{fig:0946+006} shows pronounced activity during the first $\sim$8 years of
the mission, and lower level activity until MJD 59000,
but relative quiescence thereafter.
High cadence radio monitoring over the first years of the
{\it Fermi} mission have been described by
\cite{Foschini2012}, \cite{Angelakis2015} and \cite{Lahteenmaki2017},
with the frequent OVRO monitoring at 15\,GHz in particular
revealing a series of peaks separated by several months.
The ATCA monitoring suggests a peak around MJD 55550,
which is confirmed by the works just cited.
The highest 9.0\,GHz flux density in the ATCA monitoring
at MJD 56450
coincides with a peak seen between 8 and 23\,GHz by \cite{Angelakis2015}.
The next highest 9.0\,GHz flux densities occur around
MJD 57800, while the source was somewhat less active
at gamma-rays.
Most notable, however, is a pronounced (and better-sampled)
radio flare starting around MJD 59200, which shows the
"classical" radio flare characteristics of peaking earlier
and higher at higher frequencies, with some evidence for
a small secondary peak around MJD 60000. 
This radio flare
however is not accompanied by any significant gamma-ray activity.

\subsubsection{PKS\,1244$-$255} \label{sec:1244-255}

\begin{figure*}
	\includegraphics[width=2\columnwidth]{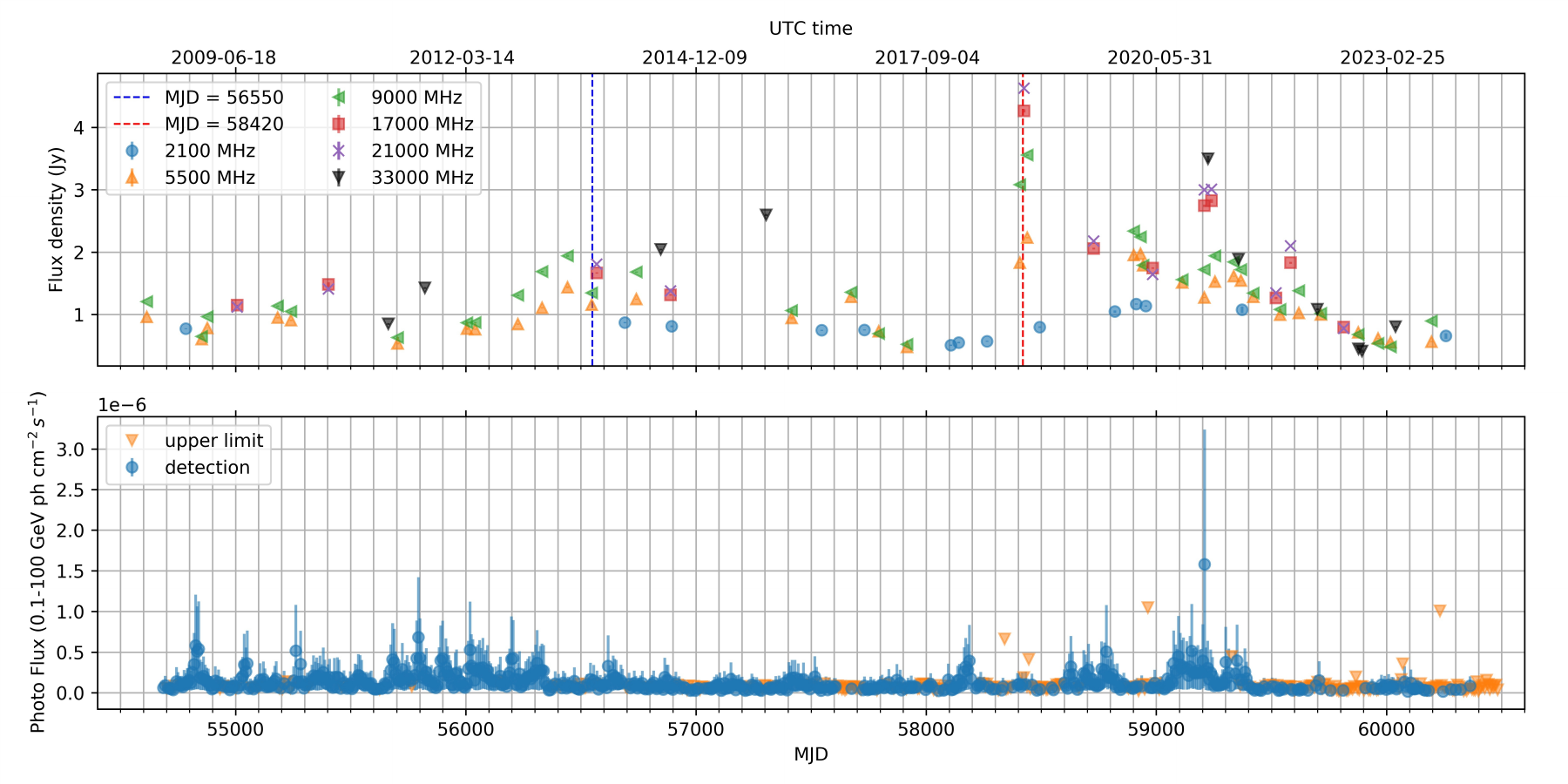}
    \caption{The radio and gamma-ray light curves of PKS 1244$-$255. The dashed lines indicate the epochs for which radio spectra are shown in Figure~\ref{fig:spectra}.}
    \label{fig:1244-255}
\end{figure*}

The {\it Fermi} light curve for PKS\,1244$-$255
in Fig.~\ref{fig:1244-255} shows a prolonged period of activity
from MJD 54700 to MJD 57000, followed by shorter flares
around MJDs 58200, 58700 and 59200.
The radio monitoring is sparse, but there is a clear
doubling in 9.0\,GHz flux density between MJD 56000
and MJD 56400, accompanied by a lagging increase in the 5.5 GHz
flux densities.
(We have excluded one epoch of 7\,mm data near MJD 56120,
for which the pipelined data analysis had yielded a
value above 4\,Jy. Inspection of other sources observed
close in time at that epoch revealed other unexpectedly
high values, and so we have omitted that data point here.)
The highest flux densities over the 15 year period occur
near MJD 58400, after the MJD 58200 gamma-ray peak.
The radio monitoring is particularly infrequent around this time,
but the two epochs near MJD 58400 of ATCA monitoring confirm this radio high state.
Since that time, the radio flux densities have generally been in a long-term decline, though with an increase coinciding with the MJD 59200 gamma-ray activity.
In recent years the gamma-ray activity has been at its lowest level.

\subsubsection{PKS\,1502+036} \label{sec:1502+036}

\begin{figure*}
        \includegraphics[width=2\columnwidth]{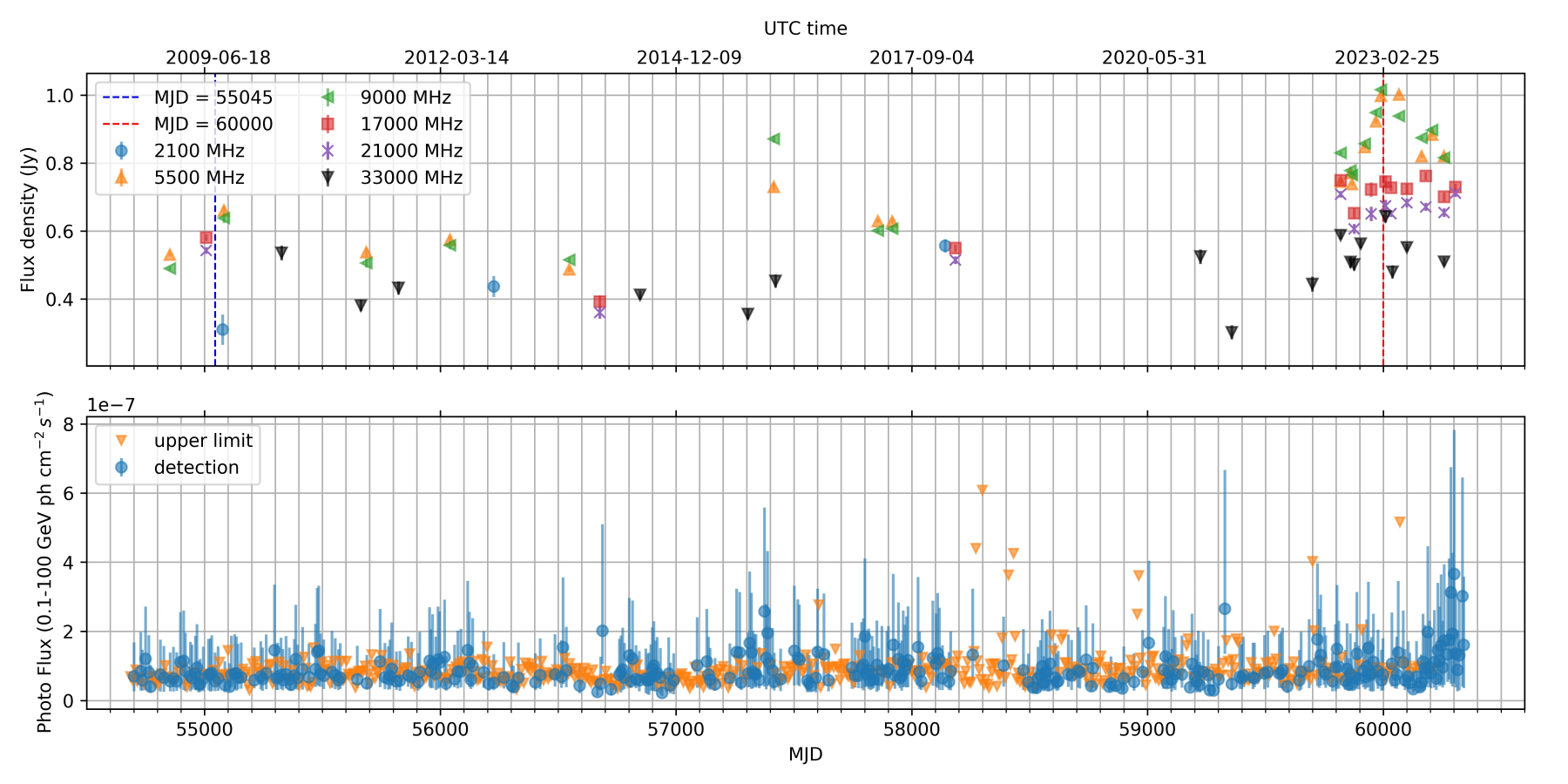}
    \caption{The radio and gamma-ray light curves of PKS 1502+036. The dashed lines indicate the epochs for which radio spectra are shown in Figure~\ref{fig:spectra}.}
    \label{fig:1502+036}
\end{figure*}

The {\it Fermi} light curve of PKS\,1502+036 in Fig.~\ref{fig:1502+036} indicates it is the
faintest of the sources considered here.
Prior to MJD 59600, the sparse radio data peaked
around MJD 57400.
Somewhat surprisingly, the radio data post MJD 59600
shows a clear peak at 5.5 and 9.0\,GHz with
the flux densities at 17, 21 and 33 GHz staying almost constant.
The 5.5 and 9.0\,GHz peak at MJD 60000 comes near
the start of the most significant gamma-ray activity
in this source.

\subsubsection{PKS\,2004$-$447} \label{sec:2004-447}

\begin{figure*}
	\includegraphics[width=2\columnwidth]{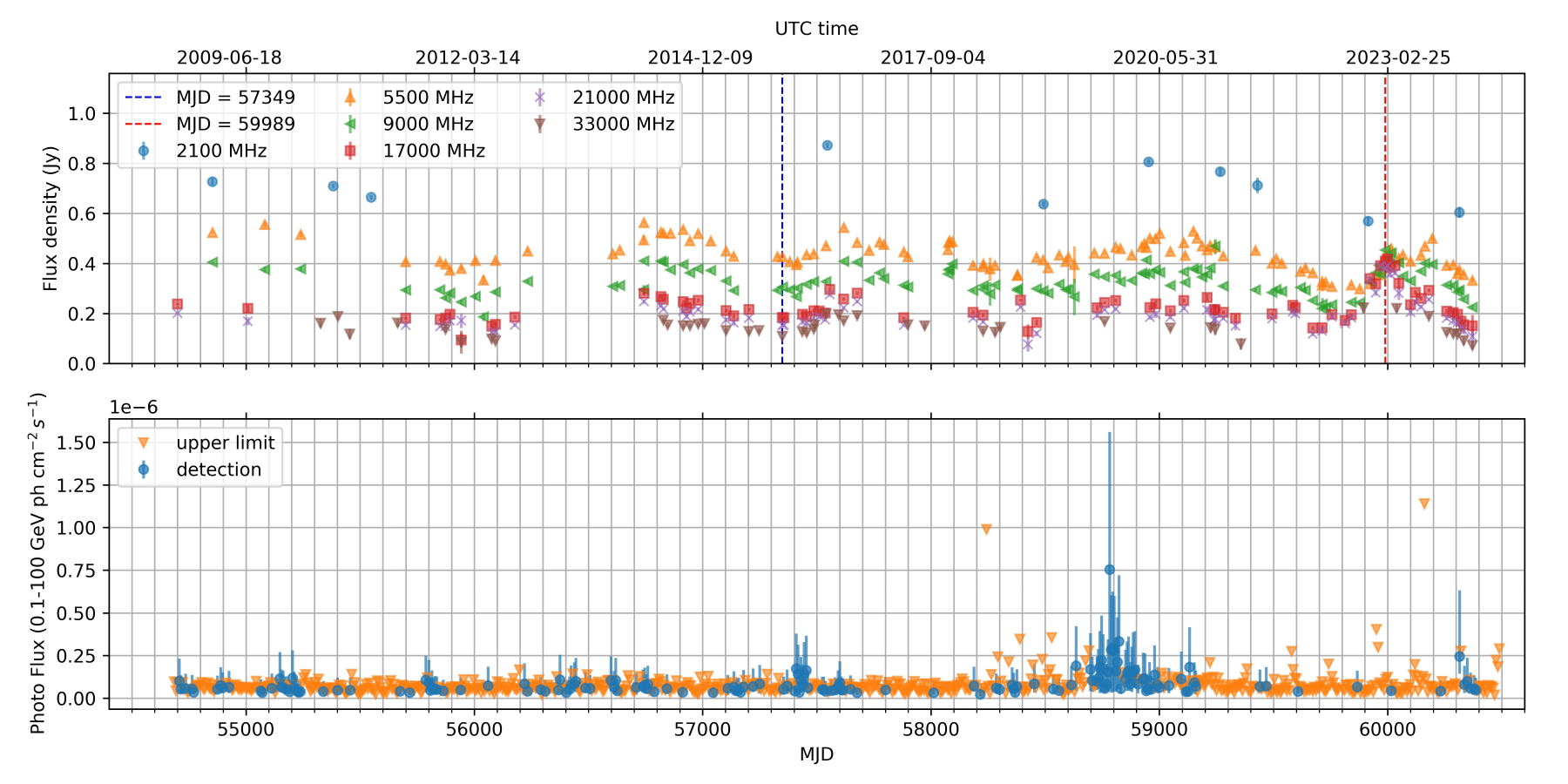}
        \includegraphics[width=2\columnwidth]{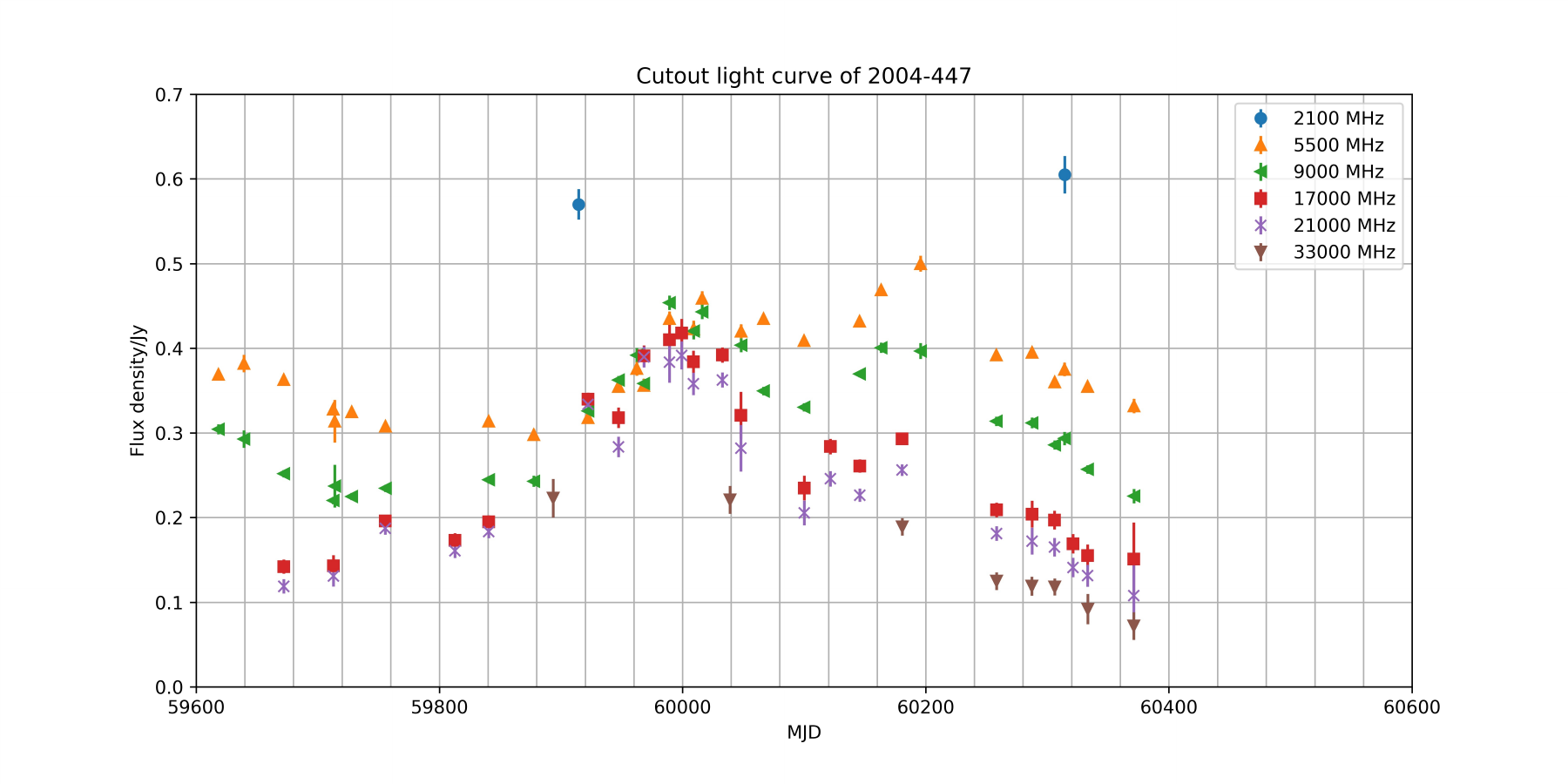}
    \caption{The radio and gamma-ray light curves of PKS 2004$-$447. The dashed lines indicate the epochs for which radio spectra are shown in Figure~\ref{fig:spectra}. The bottom panel is an expanded view of the radio flare starting around MJD 59900.}
    \label{fig:2004-447}
\end{figure*}

Although considered as a NLS1 previously \citep{Oshlack2001},
\cite{Foschini2022} label PKS\,2004$-$447 a misaligned jetted AGN.
\cite{Berton2021} confirm that the optical spectrum
meets the classification of an NLS1, but propose that the source is a hybrid
Compact Steep Spectrum (CSS)/NLS1, based on its radio properties, noting that
these hybrid characteristics are similar to 3C\,286.
The {\it Fermi} light curve in Fig.~\ref{fig:2004-447} reveals that the source is largely quiescent, being detected in only 20\% of
one-week bins, with the exception being a prominent gamma-ray flare in 2019. 
\cite{2021A&A...649A..77G}
made a multi-wavelength examination of data around this time,
and report significant increases in optical/UV fluxes coincident
with the flare. The gamma-ray flare occurred during a several-year period of gradually
increasing radio flux densities, but did not result in any
dramatic short-term radio outburst.
More recently, as revealed by Fig.~\ref{fig:2004-447}, there has been a pronounced
radio flare, with frequencies between 5.5 and 21\,GHz increasing
simultaneously over $\sim$6 months from MJD 59840.
The higher frequencies then decline more rapidly than lower frequencies,
with a secondary radio peak around MJD 60200.
These are the highest flux densities over the 10 years in the 1\,cm band, however, the radio flare is not accompanied by any appreciable activity at gamma-ray energies.

\subsection{Variability indices}
We follow \cite{Tingay2003} in calculating a radio variability index, defined
as the rms variation about the mean, divided by the mean flux density,
in each band for each source. \cite{Tingay2003} studied 185 radio-loud (5\,GHz flux density $>\sim$1\,Jy) AGN over a 3.5-year period in the 1.4, 2.5, 4.8 and 8.4\,GHz bands. That study yielded median variability indices in those bands of 0.06, 0.06, 0.08 and 0.09 respectively, with 90\% of sources having variability indices
below 0.20, 0.20, 0.24 and 0.28.
It is apparent 
from Table~\ref{tab:gamma_variability_factor}
that the five gamma-ray loud NLS1 sources we have studied all have variability indices generally well in excess of these median values, and PMN J0948+0022 and PKS 1244$-$255 have variabilities in the top 10\% of the \citet{Tingay2003} sample. This is consistent with the finding of \citet{Tingay2003} that gamma-ray loud AGN (i.e., at that time, EGRET-detected AGN) tended to have higher variability indices than gamma-ray quiet AGN.

\subsection{Spectral variation}
\begin{figure*}
    \includegraphics[width=1\columnwidth]{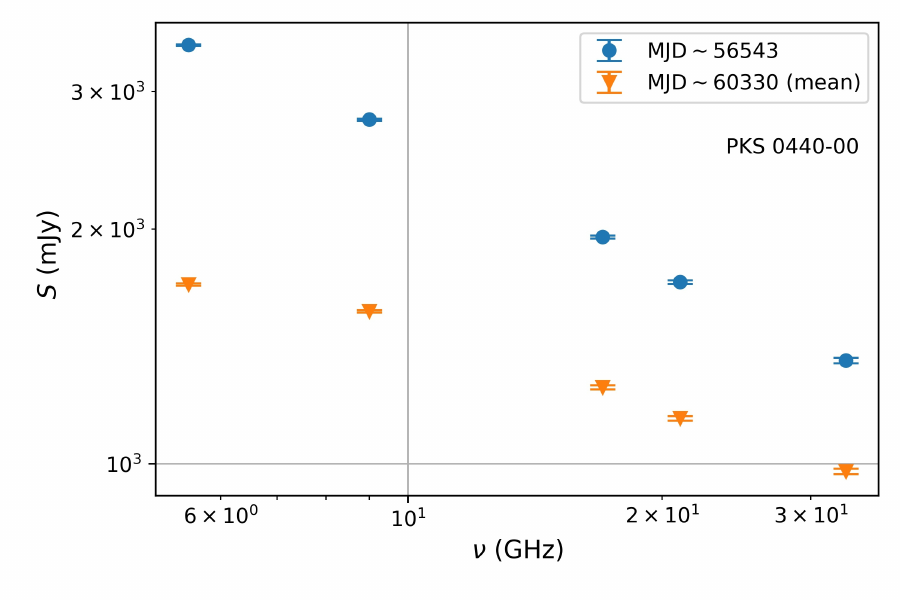}[a]    
    \includegraphics[width=1\columnwidth]{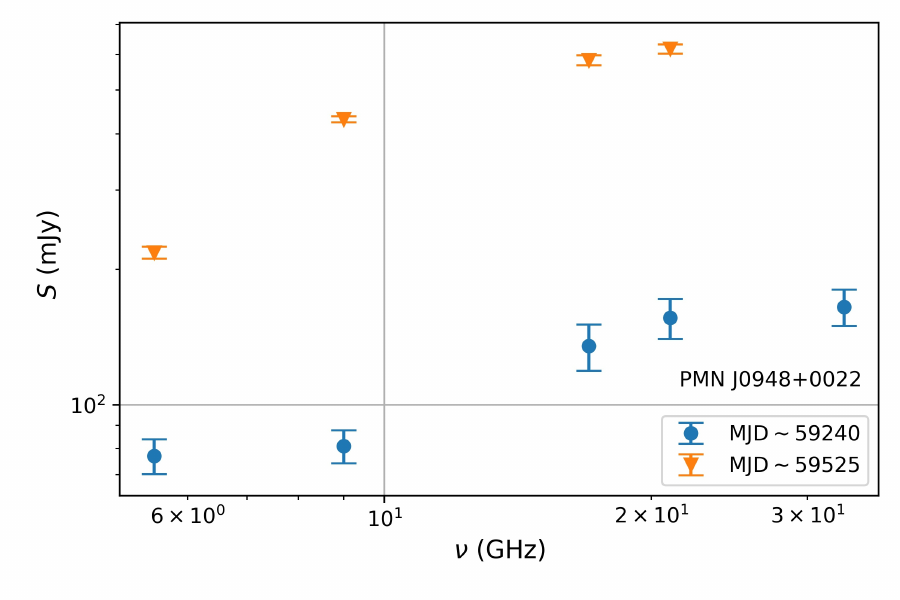}[b]
    \includegraphics[width=1\columnwidth]{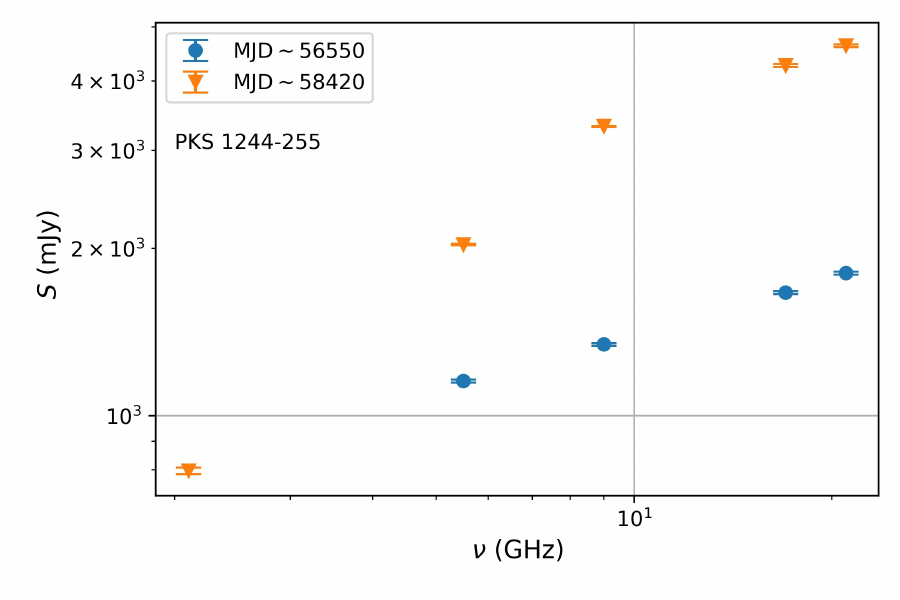}[c]
    \includegraphics[width=1\columnwidth]{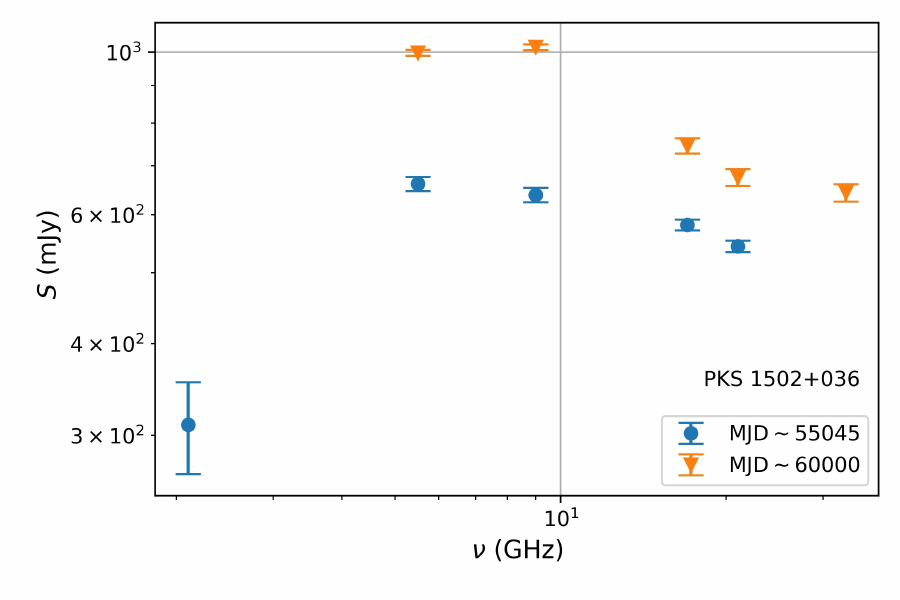}[d]
    \includegraphics[width=1\columnwidth]{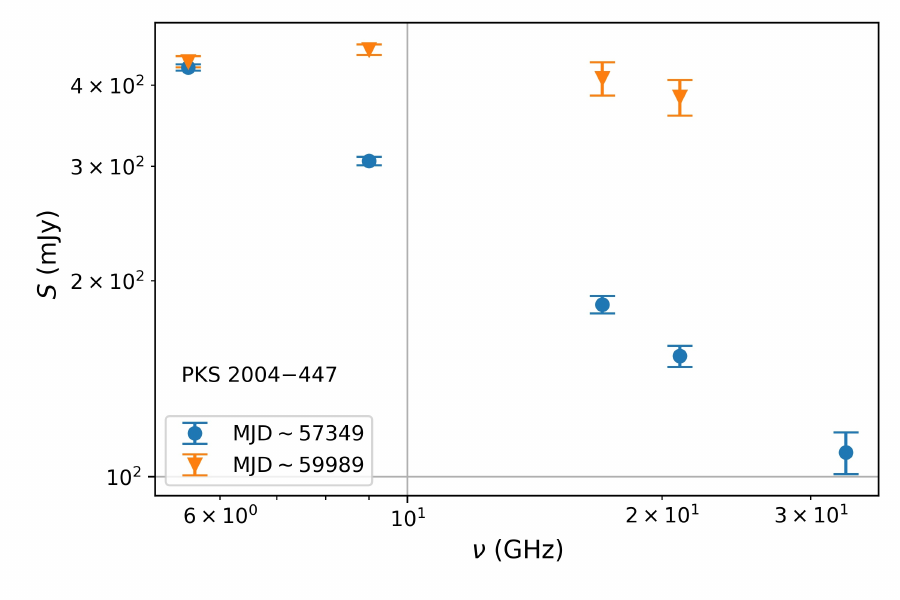}[e]
    \includegraphics[width=1\columnwidth]{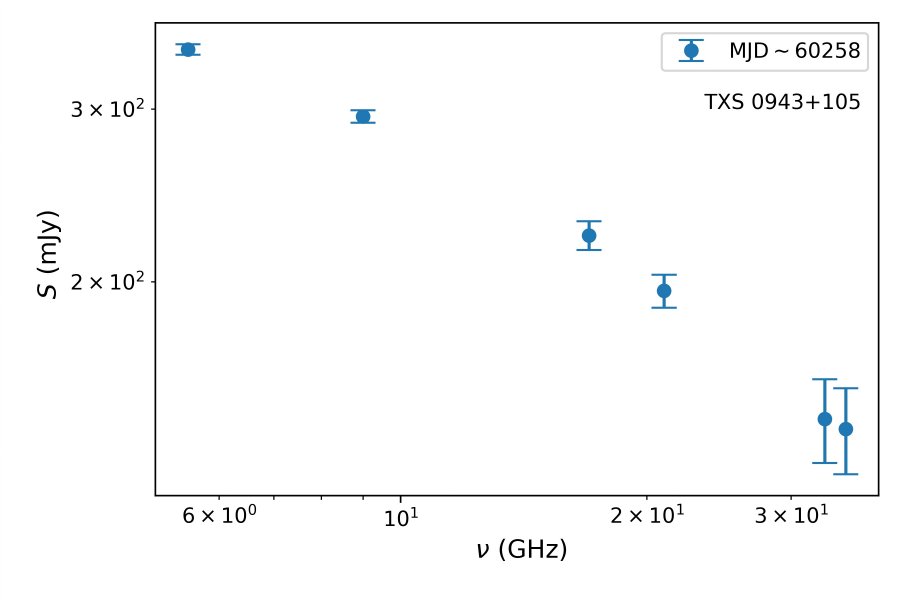}[f]
    \caption{Panel [a] to [e] are the spectra for the two epochs indicated in Figures 1 to 5 for PKS 0440$-$00, PMN J0948+0022, PKS 1244$-$255, PKS 1502+036, and PKS 2004$-$447, respectively. Panel [f] is the single epoch spectrum for TXS\,0943+105.}
    \label{fig:spectra}
\end{figure*}

A detailed wide-band study of the variation in spectra in the preceding radio light curves is complicated by the fact there are only few epochs with simultaneous broad-band data points. Nevertheless, there are enough epochs with observations in at least four bands to make an examination of the range of variations seen in spectra. In Fig.~\ref{fig:spectra}, we show spectra at two epochs for each of the five sources considered in the preceding sections. The two epochs have been selected in an attempt to show the extremes of the spectral indices. The values of the spectral indices are tabulated in Table~\ref{tab:gamma_inband_index}.
Several features are apparent: 
while several spectra can be reasonably approximated by a single spectral index
over a wide range, more often there is a range of spectral indices present at a single epoch. The difference between flattest and steepest spectral indices in this comparison of two epochs for a given source ranges from 0.49 for PKS 0440$-$00 to 1.28 for PMN J0948+0022 and PKS 1502+036.
Not surprisingly, this range of variation in spectral index results in some sources having steep ($\alpha \leq -0.5$) spectra at some epochs but flat spectra at others. 
Clearly, for variable sources such as these, care must be taken in obtaining contemporaneous flux densities for any meaningful spectral index studies.

Additionally in Fig.~\ref{fig:spectra} we show a single epoch
spectrum between 5.5 and 34 GHz for the candidate NLS1 source TXS 0943+105 (considered as a FSRQ by \cite{Foschini2022}) for comparison with future observations. Examination of the {\it Fermi}-LAT LCR indicates this source was in a less active gamma-ray state
at this epoch, with fewer detections on one-week timescales, compared to the first $\sim$8 years of the {\it Fermi} mission.

\section{Gamma-ray quiet NLS1 galaxies} \label{sec:gamma-quiet}

\cite{2018A&A...615A.167C} presented a catalogue of  Southern Hemisphere NLS1s derived from the Six-degree Field Galaxy Survey (6dFGS) final data release, classifying 167 NLS1s by their optical spectral properties. 
\cite{2020MNRAS.498.1278C} used the VLA in C-configuration to observe 62 of these sources with declinations $> -25^\circ$ at 5\,GHz with an image sensitivity of $\sim 7 \mu{\rm Jy}$. 
Subsequently, \cite{2022MNRAS.512..471C} observed 85 of these sources (with declination $< -25^\circ$) at 5.5 and 9.0\,GHz with the ATCA. 
As the VLA is more sensitive, and the observation times were similar, the VLA observations had a higher detection rate than the ATCA observations, as described in Section\,\ref{sec:intro}.

As a pilot survey for a potential larger program,
we selected 21 sources from the sample of \cite{2018A&A...615A.167C} for observations in ATCA Director's Discretionary Time under the observing code CX540. These sources are listed in Table~\ref{tab:flux_spectral_index}.
Ten of the sources had been observed by \cite{2020MNRAS.498.1278C}
and five by \cite{2022MNRAS.512..471C}, 
and our observations were intended to determine whether these sources showed evidence of variability at radio frequencies.
The other six sources were among 21 
sources which were not able to be observed as part of the \citet{2020MNRAS.498.1278C}
VLA sample.

We also collect data from the Rapid ASKAP Continuum Surveys
at 887.5 MHz \cite[RACS-Low]{2021PASA...38...58H} and 1367.5 MHz \cite[RACS-Mid]{2023PASA...40...34D} for our sample.  
RACS-Low covers the south of a declination of +41 degrees, with an median RMS noise level in images of 0.25 mJy/beam. 
RACS-Mid covers the sky south of a declination of +49 degrees, with Stokes~I images having a median noise level of 0.20 mJy/beam.

Due to the low flux density of our targets, observations (using project code CX540) were made with 10$\sim$15 minute snapshots bracketed by 2-minute scans on a nearby phase calibrator. The ATCA primary flux density calibrator, PKS 1934$-$638, was also used for bandpass calibration. 
Data reduction was carried out in MIRIAD \citep{1995ASPC...77..433S} 
using standard procedures.
During the calibration, radio frequency interference (RFI) was flagged. 
Flux densities were measured using the miriad task "imfit".

\subsection{Results}\label{sec:res}

Of our sample of 21 sources, we detected 14 sources with the ATCA at 5.5\,GHz, with 10 sources also detected at 9.0\,GHz. 
We measured the integrated flux densities of the detected sources and set upper limits of three times the rms noise level for the non-detections. 
Note again that the reported errors are statistical only: systematic errors of up to 5\% may also be present \citep[see, e.g.][]{Tingay2003}.
Fifteen of the 21 sources were matched in
RACS-Low, and 16 in RACS-Mid.  

With these flux densities we can examine the spectra of these sources. 
We calculate the spectral index, $\alpha$, between bands as,
\begin{equation}
    \alpha= \frac{\log (S_{\nu_1}/S_{\nu_2})}{\log (\nu_1/\nu_2)},
\end{equation}\label{eq:alpha}
where $S_{\nu_1}$ and $S_{\nu_2}$ are the flux densities at frequencies $\nu_1$ and $\nu_2$, respectively. The error in $\alpha$ is derived following the propagation of errors. 
We follow the usual convention of categorizing the spectrum 
as "steep" if $\alpha \leq -0.5$ and "flat" if $\alpha > -0.5$.
Fourteen of the 15 sources detected in both RACS surveys
have steep spectra between 887.5 and 1367.5 MHz (given as $\alpha_1$ in Table~\ref{tab:flux_spectral_index}).
Between 1367.5\,MHz and 5.5\,GHz,
twelve sources have a steep spectrum and 
four targets show a flat spectra ($\alpha_2$).
Finally, between 5.5 and 9.0\,GHz,
eight sources have steep spectra while six sources have flat spectra ($\alpha_3$). 

\begin{figure*}
    \includegraphics[width=2\columnwidth]{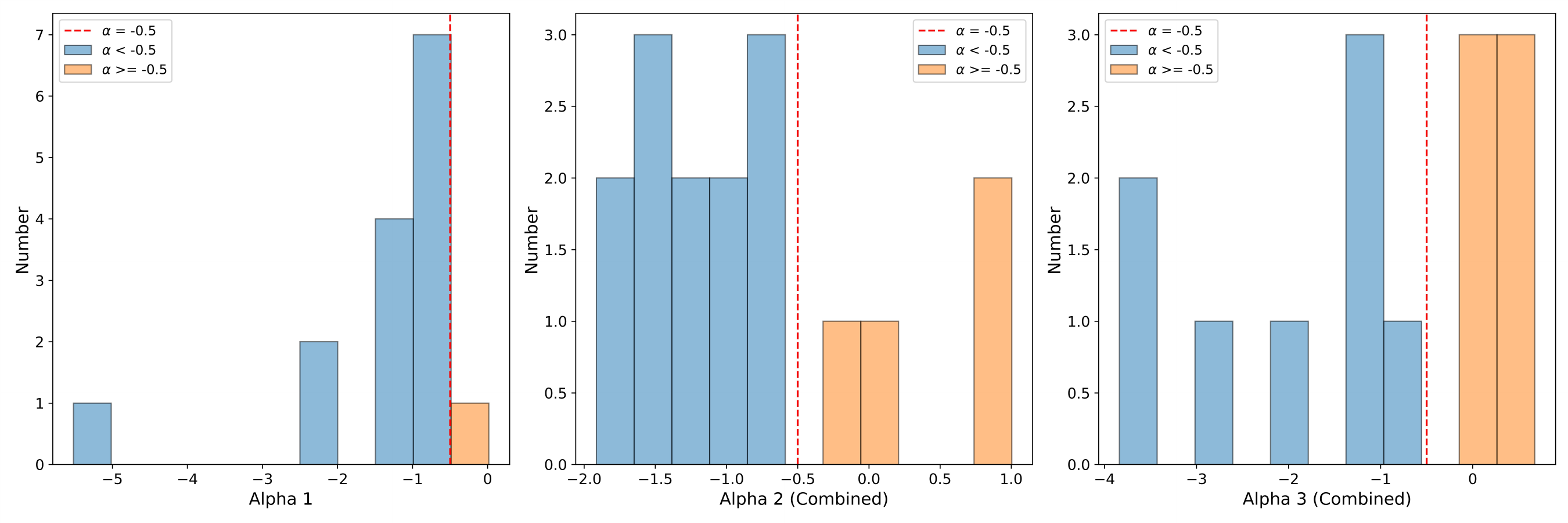}
    \caption{The histogram of the spectral indices for gamma-ray quiet sources in the \protect\cite{2020MNRAS.498.1278C,2022MNRAS.512..471C} sample. The left panel ($\alpha 1$) is the spectral index between RACS-Low (887.5\,MHz) and RACS-Mid (1367.5\,MHz). The central panel ($\alpha 2$) is the spectral index between RACS-Mid and 5.5\,GHz. The right panel ($\alpha 3$) is the spectral index between 5.5 and 9.0\,GHz (the observation of J1515$-$7820 on 2024 Feb 17 was selected) and the "combined" in x-axis presents the data including the upper limits on the spectral index.}
    \label{fig:histogram}
\end{figure*}

The medians of $\alpha_1$, $\alpha_2$ and $\alpha_3$ are $-1.0\pm0.5$, $-0.9\pm-0.0$ and $0.0\pm0.3$, respectively.
Details of spectral indices are displayed in Table~\ref{tab:flux_spectral_index} and the overall distribution of the spectral index types of our sample is depicted in Fig.~\ref{fig:histogram} including the sources with upper limits showing obvious spectral type. 
Note that we observed two sources, J1500$-$7248 and J1515$-$7820, 
twice: we take the data observed on 2024 February 17  for the spectral index calculation due to their longer integration times. The potential mechanisms result in various spectral index will be discussed in Section~\ref{discussion}.

\subsubsection{Sources previously observed with the VLA}
We observed ten sources selected from \citet{2020MNRAS.498.1278C} which were detected by VLA at 5.5\,GHz. 
All ten sources were detected in both RACS-Low and RACS-Mid.
We detected six of these sources at both 5.5 and 9.0\,GHz, 
and three at only 5.5\,GHz.
As expected, the ATCA non-detections were generally the faintest of the \cite{2020MNRAS.498.1278C} detections.
The main reason for re-examining these sources was to search for evidence of radio variability among these gamma-ray quiet NLS1s, and several sources have varied significantly over the $\sim$5 years between the VLA observations and our ATCA observations,
most notably J0122$-$2646 (from 0.9 to 7.1\,mJy), J0452$-$2953 (from 3.4 to 0.7\,mJy) and J0447$-$0508 (from 4.0 to 2.4\,mJy).
(We note that for J0447$-$0508, Table~\ref{tab:flux_spectral_index}
lists integrated flux densities of 89.0\,mJy for RACS-Low yet only 8.2\,mJy for RACS-Mid. RACS-Mid resolves this source into two distinct components, yet with the lower angular resolution of RACS-Low, they appear as a single extended source.) 

\subsubsection{Sources previously observed with the ATCA}
Five sources had previously been observed with the ATCA by \cite{2022MNRAS.512..471C}.  The brighter sources show strong
variability: J1057$-$4089 (from 168 to 283 mJy at 5.5\,GHz and from 152 to 301 mJy at 9.0\,GHz) and J1500$-$7248 (from 31 mJy to 54 and 79 mJy at 5.5\,GHz and from 35 mJy to 60 mJy at 9.0\,GHz in the two epochs we observed it).
We note that \cite{2022MNRAS.512..471C} find these two sources 
both show significant structure, which could potentially result in flux density not being fully recovered in short snapshot observations (if the position angle of the synthesized differs significantly from that of the source structure).  However, the \cite{2022MNRAS.512..471C} observations were made in a 750m array, whereas our observations, which yielded higher flux densities, were made in 6\,km arrays, suggesting resolution effects are unlikely to be a significant factor in this case.

J0609$-$5606 was a faint detection previously (0.4\,mJy) but was not re-detected. J0307$-$7250 was of particular interest as it had previously been detected at 9.0\,GHz but not 5.5\,GHz. We did not detect it at either frequency, and note further that the source is not listed in either of the RACS catalogues.
However, our upper limit at 9.0\,GHz is only just below the reported flux density of \cite{2022MNRAS.512..471C}. At face value, these facts suggest a (possibly variable) inverted spectrum source,
which merits further follow-up. J1515$-$7820 is the only source that shows different spectral types in 4\,cm band in two slots, as the flux density at 9.0\,GHz significantly decreased while that at 5.5\,GHz remained the same.

\subsubsection{Sources not previously observed}
As \cite{2020MNRAS.498.1278C} were not able to observe all sources with declinations $>-25^\circ$, there are 13 sources in the original observing catalogue \citep{2018A&A...615A.167C} that are not considered in the final paper. We observed six of these sources (J0133$-$2109, J2137$-$1112, J2229$-$1401, J2244$-$1822, J2250$-$1152 and J2311$-$2022)
for relatively short integrations, mostly 20 minutes. One source, J2250$-$1152, was detected at both 5.5 and 9.0\,GHz, and another, J2229$-$1401, was detected at 5.5\,GHz only. 
It is noteworthy that neither of these sources were detected in the RACS-Low or RACS-Mid surveys, despite simple extrapolations from the ATCA detections suggesting the RACS-band flux densities would be well above the respective noise levels. This suggests that either the spectra are more complex, or that the sources have varied significant over the intervening two years, or both.

\section{Discussion}{\label{discussion}}
\subsection{The variabilities}

Archival multi-frequency ATCA radio data has allowed the radio state of five gamma-ray loud NLS1s to be compared to their gamma-ray state, as indicated by data from the {\it Fermi}-LAT LCR. 
The {\it Fermi} LCR data reveals a wide range in activity levels of 
detected NLS1s: PKS 1244$-$255 was detected in 73\% of the weekly-binned data, whereas PKS 0221+067 and GB6 J1102+5249 were only detected 6\% of the time (Table~\ref{tab:fermi_NLS1s}).
The radio variability indices for the five sources we considered are generally well above the median variabilities determined for a sample of radio-loud AGN by \cite{Tingay2003}. 
Although the cadence of our radio data is limited, particularly for the early years of the {\it Fermi} mission, we are still able to draw a number of conclusions. 
There are peaks in our radio light curves that are consistent with those previously reported
\citep{Foschini2012,Angelakis2015,Lahteenmaki2017}.
In some cases radio flares are contemporaneous with gamma-ray outbursts, e.g., PKS 0440$-$003 in 2014, PMN J0948+0022 in 2010 and 2013 and PKS 1244$-$255 in 2021. 

The sparsity of the radio monitoring, and the fact that gamma-ray active states can continue for many years, preclude any conclusions on time lags between gamma-ray and radio activity. However, it is evident that radio outbursts are not necessarily proportionate to gamma-ray activity: the under-sampled 2018 radio flare in PKS 1244$-$255 follows a small increase in gamma-ray flux, whereas five years of similar gamma-ray activity at the start of the {\it Fermi} mission coincided with more modest radio variability. 

There are also several examples of pronounced radio outbursts which are not accompanied by any perceptible change in gamma-ray state, e.g., the 2021 flare in PMN J0048+0022, and the 2022 flare in PKS 2004$-$447.
Similar behaviour has been reported for other sources \citep[e.g.][]{2003ApJ...590...95L}.
Perhaps most surprising is the 2022 flare seen at 5.5 and 9.0\,GHz for PKS 1502+036, which does not appear to have a counterpart at higher radio frequencies.  \cite{Angelakis2015} note the occurrence of radio flares which "disappear" below some frequency in the cm-band, but we are not aware of flares disappearing above some cutoff. 

\cite{Foschini2012} note
the variability of gamma-ray and radio fluxes 
from PMN J0948+0022 in 2009 and 2010 followed the canonical expectations of relativistic jets, with gamma-rays leading the radio emission by a few months, but that the multi-frequency behaviour was less clear in 2011.
In particular, they note an "orphan" optical/X-ray flare in October 2011 with no counterpart at gamma-ray energies.
\cite{2014MNRAS.438.3521D} also examine the 
multi-wavelength behaviour of PMN J0948+0022
in 2011, noting that variability observed in optical and near-infrared in April/May has no counterpart at gamma-ray energies. 
They suggest this behaviour could be related to a bending and inhomogeneous jet, possibly containing a turbulent multi-cell structure.

A key goal for our study of a gamma-ray quiet sample
of NLS1s was to determine whether this class also
showed evidence for variability at 5.5\,GHz. 
This class of sources also tends to be much fainter
at radio wavelengths, with longer integrations required
to detect (sub-)mJy--level sources. 
Comparison of our measured flux densities with
those of \cite{2020MNRAS.498.1278C} and \cite{2022MNRAS.512..471C} revealed clear evidence of variability ($>3\sigma$) for over half of the 14 sources detected at two epochs over the several year period between observations. (And there are another three undetected sources whose variations between two epochs exceed $3\sigma$ in their statistical errors.)
We acknowledge that the effects of differing angular resolutions, incomplete ($u,v$) coverages, and different primary flux density calibrators between the VLA and ATCA may contribute to apparent variability, for the fainter sources in particular, however it is incontrovertible that a number of
these gamma-ray quiet NLS1s show significant radio variability.
At face value, this might be seen as evidence in favour of jetted radio emission over a predominantly star formation (SF) origin in these variable sources, however a more definitive test would be with follow-up VLBI observations of these sources
\citep[cf][]{2015ApJS..221....3G,2018Lister,2023ApJ...943..136S}.

\subsection{The spectral behaviour}

For the five gamma-ray loud NLS1s for which we have multi-epoch multi-frequency data, we were able to compare the spectral behaviour at two contrasting epochs.
These confirmed that these gamma-ray and radio-loud sources tend to have flat or inverted radio spectra at most epochs, though the actual value of the spectral index can vary significantly.
This highlights that contemporaneous flux density measurements are required for any meaningful spectral index studies
of such variable sources.
There are a few cases where the flux densities exceed 2\,Jy at multiple frequencies (with inverted spectra), such as PMN J0948+0022 and PKS 1244$-$255, indicating that the dominant contribution is
from the relativistic jet (especially for PMN J0948+0022, which displays a parsec-scale jet morphology) rather than SF.

In contrast, the gamma-ray quiet sample tend to have steep spectra. We combined RACS-Low (887.5 MHz) and RACS-Mid (1367.5 MHz) data with our 5.5 and 9 GHz data, to determine up to three spectral indices. A number of sources had steep ($\alpha < -0.5$) spectra across this whole frequency range.
No source had flat spectra between all frequencies, though a number of sources had flat spectra at the higher frequencies,
where the median spectral index was 0.0$\pm$0.3 (see Table~\ref{tab:flux_spectral_index}
there was a tendency for the more variable sources to have flatter spectra, but also exceptions to this trend (such as J1044$-$1826)).

Spectra, like variability, can also provide clues as to 
the origin of the radio emission.
Radio spectra can be made up of contributions from SF, relativistic jets, accretion disk corona, AGN-driven winds or outflows, free-free emission (FFE) and even external Compton processes. Diffuse emission is widely observed in the SF radio morphology on the host galaxy scale with optically thin spectra from several hundred MHz to $\sim$10 GHz following $\alpha = -0.8\pm0.4$ \citep{2019ApJ...875...80G,2021MNRAS.507.2643A} without apparent spectral breaks or turnover.
Jet and wind contributions are difficult to distinguish based on the spectra and morphology, as both can launch pc to kpc linear structures \citep{2018A&A...614A..87B,2020MNRAS.498.1278C,2021MNRAS.508.1305Y} but the spectra would be expected to be steeper if extended emission dominates, or flatter if a compact core dominates. 
For radio emission from the corona, the morphology would be compact and optically thick on sub-pc scales, requiring VLBI observations to determine whether the jet base has high brightness temperature and whether the radio emission is collimated or not. 
A spectral index of $-0.1$ would be expected if FFE dominates. Of our sample, only J1057$-$4039 approached this value,
with $\alpha_2=-0.1\pm0.0$.
Three of our sources might have spectra close to expectations for a SF dominated case: J0122$-$2646 has $\alpha_1=-0.8\pm0.8$, J0400$-$2500 has $\alpha_2=-0.8\pm0.2$ and J0447$-$0508 has $\alpha_3=-0.8 \pm 0.4$.

However, these consistencies in only specific frequency intervals can't be taken as the conclusive evidence of the emission mechanism, as the models predict spectral indices over broader frequency ranges. Therefore, the emission mechanism in our gamma-ray quiet NLS1s remains unclear, with further studies required.

\section{Conclusion}{\label{conclusion}}

In this paper we have studied two classes of NLS1 galaxies.
For a sample of five gamma-ray loud NLS1s, 
we have compared the $\sim16$ year gamma-ray light curves from \textit{Fermi} with archival
multi-frequency radio monitoring with the ATCA between 2.1 and 33 GHz. Although the radio data is sparse, particular in the early years, a number of conclusions can be drawn.

\begin{itemize}
    \item These variable gamma-ray sources are also highly variable radio sources.
    \item Radio flares can in some instances be associated with gamma-ray high states.
    \item There are also significant radio flares with no corresponding increase in gamma-ray activity.
\end{itemize}

We have extended the study of candidate gamma-ray quiet NLS1 galaxies from the sample of \cite{2018A&A...615A.167C} with new ATCA observation and added ASKAP survey data as well.
As a pilot survey for a potential larger program,
we selected 21 sources, 
ten of which had been detected with the VLA
by \cite{2020MNRAS.498.1278C},
five of which had been detected with the ATCA by \cite{2022MNRAS.512..471C}, 
and six of which had not been previously observed at 5.5\,GHz.
We compiled data from the ASKAP RACS-Low and RACS-Mid surveys  around 1 GHz for these sources and calculated the spectral 
indices.

\begin{itemize}
    \item About half of the 14 sources detected at two epochs showed evidence of radio variability.
    \item Although the spectra between 887.5 MHz and 9.0\,GHz were generally steep, a number of sources, and preferentially the variable sources, had flat spectra at the higher frequencies (1367.5\,MHz--5.5\,GHz and 5.5--9.0\,GHz).
    \item Two of six sources previously not observed at 5.5\,GHz were detected for the first time. 
\end{itemize}

While illuminating, the under-sampled nature of the archival ATCA light curves results in our comparison with gamma-ray light curves being qualitative rather than quantitative. 
As stressed by other workers \citep[e.g.,][]{Foschini2012,2023MNRAS.523..441Y} higher cadence multi-frequency monitoring is required to obtain data sets amenable to detailed SED modelling. We plan to continue higher cadence multi-frequency monitoring of the gamma-ray loud sources with the ATCA.

\begin{table*}
\caption{{\it Fermi} detected NLS1s based on \citet{Foschini2022}.}
\label{tab:fermi_NLS1s}
\begin{tabular}{llcllccccc} 
\hline
Catalogue name	&	$Fermi$ Name	&	$z$	&	$\rm Class_{\rm original}$	&	 $\rm Class_{\rm revised}$	&	$N_{\rm detection}$	&	$N_{\rm~upper\,limit}$	&	$R_{\rm detection}$	&	$ Ph_{\rm~highest}$		&	$Ave(F_{\rm 5.5\,GHz})$ 	\\
(1)  & (2) & (3) & (4) & (5) & (6) & (7) & (8) & (9) & (10)\\
\hline
1H 0323+342                 	&	J0324.8+3412 	&	0.063	&	nlsy1	&	 NLS1 	&	203	&	612	&	0.25 	&	9.58E-07	&	0.350 	$\pm$	0.005 	\\
SBS 0846+513                	&	J0850.0+5108 	&	0.584	&	NLSY1	&	 NLS1 	&	177	&	644	&	0.22 	&	4.19E-07	&				\\
PMN J0948+0022              	&	J0948.9+0022 	&	0.584	&	NLSY1	&	 NLS1 	&	423	&	380	&	0.53 	&	7.34E-07	&	0.266 	$\pm$	0.001 	\\
IERS B1303+515              	&	J1305.3+5118 	&	0.785	&	nlsy1	&	 NLS1 	&		&		&		&		&				\\
B3 1441+476                 	&	J1443.1+4728 	&	0.703	&	nlsy1	&	 NLS1 	&		&		&		&		&				\\
PKS 1502+036                	&	J1505.0+0326 	&	0.408	&	NLSY1	&	 NLS1 	&	324	&	478	&	0.40 	&	3.66E-07	&	0.741 	$\pm$	0.002 	\\
MG2 J164443+2618            	&	J1644.9+2620 	&	0.144	&	NLSY1	&	 NLS1 	&	87	&	736	&	0.11 	&	2.98E-07	&				\\
\hline
revised class	&		&		&		&		&		&		&		&		&				\\
PKS 2004$-$447                	&	J2007.9-4432 	&	0.24	&	nlsy1	&	 MIS  	&	164	&	662	&	0.20 	&	7.54E-07	&	0.429 	$\pm$	0.001 	\\
TXS 2116$-$077                	&	J2118.8-0723c	&	0.26	&	nlsy1	&	 SEY  	&		&		&		&		&	0.103 	$\pm$	0.003 	\\
\hline
new classified	&		&		&		&		&		&		&		&		&				\\
TXS 2358+209                	&	J0001.5+2113 	&	0.439	&	fsrq 	&	 NLS1 	&	259	&	564	&	0.31 	&	9.04E-07	&				\\
GB6 J0102+4214              	&	J0102.4+4214 	&	0.874	&	fsrq 	&	 NLS1 	&	100	&	720	&	0.12 	&	1.59E-07	&				\\
PKS 0221+067                	&	J0224.2+0700 	&	0.511	&	fsrq 	&	 NLS1 	&	52	&	773	&	0.06 	&	2.80E-07	&				\\
PKS 0440$-$00                 	&	J0442.6-0017 	&	0.844	&	fsrq 	&	 NLS1 	&	464	&	361	&	0.56 	&	1.47E-06	&	2.449 	$\pm$	0.005 	\\
S4 0929+53                  	&	J0932.6+5306 	&	0.597	&	fsrq 	&	 NLS1 	&	108	&	712	&	0.13 	&	4.04E-05	&				\\
3C 232                      	&	J0958.0+3222 	&	0.531	&	fsrq 	&	 NLS1 	&		&		&		&		&				\\
PKS 1045$-$18                 	&	J1048.0-1912 	&	0.595	&	fsrq 	&	 NLS1 	&		&		&		&		&				\\
GB6 J1102+5249              	&	J1102.6+5251 	&	0.69	&	fsrq 	&	 NLS1 	&	49	&	773	&	0.06 	&	1.17E-07	&				\\
B3 1151+408                 	&	J1154.0+4037 	&	0.923	&	fsrq 	&	 NLS1 	&	91	&	730	&	0.11 	&	1.73E-07	&				\\
PKS 1200$-$051                	&	J1202.5-0528 	&	0.381	&	fsrq 	&	 NLS1 	&	119	&	706	&	0.14 	&	2.74E-07	&				\\
TXS 1206+549                	&	J1208.9+5441 	&	1.34	&	fsrq 	&	 NLS1 	&	299	&	523	&	0.36 	&	3.80E-07	&				\\
PKS B1211$-$190               	&	J1214.6-1926 	&	0.149	&	bcu  	&	 NLS1 	&		&		&		&		&				\\
PKS 1244$-$255                	&	J1246.7-2548 	&	0.638	&	fsrq 	&	 NLS1 	&	601	&	220	&	0.73 	&	1.58E-06	&	1.153 	$\pm$	0.002 	\\
TXS 1308+554                	&	J1310.9+5514 	&	0.926	&	fsrq 	&	 NLS1 	&		&		&		&		&				\\
TXS 1700+685                	&	J1700.0+6830 	&	0.301	&	fsrq 	&	 NLS1 	&	466	&	353	&	0.57 	&	1.00E-06	&				\\
MG1 J181841+0903            	&	J1818.6+0903 	&	0.354	&	fsrq 	&	 NLS1 	&	122	&	701	&	0.15 	&	1.91E-07	&				\\
B2 1846+32A                 	&	J1848.4+3217 	&	0.798	&	fsrq 	&	 NLS1 	&	210	&	613	&	0.26 	&	8.68E-07	&				\\
\hline
candidates	&		&		&		&		&		&		&		&		&				\\
GB6 J0937+5008* 	&	J0937.1+5008	&	0.28 	&	fsrq 	&	SEY	&		&		&		&		&				\\
TXS 0943+105*	&	 J0946.6+1016	&	1.00 	&	fsrq 	&	FSRQ	&	352	&	465	&	0.43 	&	2.82E-07	&	0.345 	$\pm$	0.004 	\\
OK 290	&	 J0956.7+2516 	&	0.71 	&	fsrq 	&	FSRQ	&		&		&		&		&				\\
4C +04.42*	&	 J1222.5+0414	&	0.97 	&	fsrq 	&	FSRQ	&		&		&		&		&				\\
3C 286*	&	 J1331.0+3032	&	0.85 	&	css	&	 MIS  	&		&		&		&		&				\\
TXS 1419+391	&	 J1421.1+3859	&	0.49 	&	fsrq 	&	FSRQ	&		&		&		&		&				\\
PMN J2118+0013*   	&	 J2118.0+0019	&	0.46 	&	fsrq 	&	SEY	&		&		&		&		&				\\
\hline

\multicolumn{10}{l}{$Notes$. Column (1): source names in catalogues (those marked with $^*$ are sources already known as blazars or/and $\gamma$-ray emitters); column(2): source names}\\ 
\multicolumn{10}{l}{ in $Fermi$-4FGL catalogue;  column (3): redshifts from \cite{Foschini2022}; column (4-5): original and revised classification in \citet{Foschini2022} where} \\
\multicolumn{10}{l}{ nlsy1 and NLSY1 represent firmly associated narrow-line Seyfert 1 galaxy and nlsy1 an associated narrow-line Seyfert 1 galaxy with lower confidence; NLS1} \\
\multicolumn{10}{l}{ is narrow-line Seyfert 1 galaxy, fsrq and FSRQ represent flat-spectrum radio quasar, bcu represents the unclassified source, CSS, MIS and SEY represent } \\
\multicolumn{10}{l}{compact steep-spectrum source, misaligned AGN and Seyfert galaxy, respectively; column (6-7): the numbers of $Fermi$ detections and upper limits in the }\\
\multicolumn{10}{l}{weekly-binned {\it Fermi} LAT Light Curve Repository, respectively; column (8): the detection rate calculated by $N_{\rm~detection}/(N_{\rm~detection}+N_{\rm~upper\,limit})$; column (9): }\\
\multicolumn{10}{l}{the highest photon fluxes detected by $Fermi$-LAT in the unit of $0.1-100\, {\rm GeV\,ph\,cm} ^{-2}\,s^{-1}$; column (10): the mean of flux densities at 5.5 GHz (long-term } \\
\multicolumn{10}{l}{ATCA data) in the unit of Jy.}\\
\end{tabular}
\end{table*}

\begin{table*}
\setlength\tabcolsep{4pt}
\caption{The radio variability indices of $Fermi$ detected $\gamma$-ray NLS1s.}
\label{tab:gamma_variability_factor}
\begin{tabular}{llcccccccccccc}
\hline
Catalogue name & Source &	\multicolumn{2}{c}{2100 MHz}	&	\multicolumn{2}{c}{5500 MHz}	&	\multicolumn{2}{c}{9000 MHz} &	\multicolumn{2}{c}{17000 MHz}	&	\multicolumn{2}{c}{21000 MHz}	&	\multicolumn{2}{c}{33000 MHz}	\\
&	&	N	&	$m$	&	N	&	$m$	&	N	&	$m$	&	N	&	$m$		&	N	&	$m$	&	N	&	$m$	\\
(1) & (2) & (3) & (4) & (5) & (6) & (7) & (8) & (9) & (10) & (11) & (12) & (13) & (14)\\
\hline
PKS 0440$-$00 & 0440$-$003	&	3 	&	0.08 	$\pm$	0.06 	&	9 	&	0.25 	$\pm$	0.00 	&	9 	&	0.23 	$\pm$	0.01 	&	20 	&	0.25 	$\pm$	0.01 	&	20 	&	0.24 	$\pm$	0.01 	&	17 	&	0.30 	$\pm$	0.03 	\\
PMN J0948+0022 & 0946+006	&	18 	&	0.46 	$\pm$	0.16 	&	33 	&	0.44 	$\pm$	0.03 	&	33 	&	0.50 	$\pm$	0.02 	&	12 	&	0.37 	$\pm$	0.05 	&	12 	&	0.39 	$\pm$	0.05 	&	11 	&	0.43 	$\pm$	0.03 	\\
PKS 1244$-$255 & 1244$-$255	&	14 	&	0.26 	$\pm$	0.03 	&	38 	&	0.40 	$\pm$	0.01 	&	38 	&	0.49 	$\pm$	0.01 	&	15 	&	0.49 	$\pm$	0.01 	&	15 	&	0.52 	$\pm$	0.01 	&	10 	&	0.63 	$\pm$	0.02 	\\
PKS 1502+036 & 1502+036	&	4 	&	0.29 	$\pm$	0.07 	&	18 	&	0.21 	$\pm$	0.01 	&	18 	&	0.22 	$\pm$	0.01 	&	13 	&	0.15 	$\pm$	0.02 	&	13 	&	0.16 	$\pm$	0.03 	&	17 	&	0.18 	$\pm$	0.04 	\\
PKS 2004$-$447 & 2004$-$447	&	11 	&	0.13 	$\pm$	0.02 	&	114 	&	0.14 	$\pm$	0.02 	&	114 	&	0.17 	$\pm$	0.04 	&	76 	&	0.28 	$\pm$	0.07 	&	76 	&	0.32 	$\pm$	0.08 	&	44 	&	0.24 	$\pm$	0.13 	\\
\hline
\multicolumn{14}{l}{$Notes$. Column (1) source name in catalogues; column (2) is source name in project C007; column (3, 5, 7, 9, 11, 13) are the numbers of ATCA obser-}  \\ 
\multicolumn{14}{l}{vations at each frequency, column (4, 6, 8, 10, 12, 14) are the radio variability index of each light curve,
calculated as the rms about the mean, divided by} \\
\multicolumn{14}{l}{ the mean flux density.}\\
\end{tabular}
\end{table*}

\begin{table*}
\caption{The spectral indices of $\gamma$-ray NLS1s, from the epochs shown in Figures 1 to 5.}
\label{tab:gamma_inband_index}
\begin{tabular}{llcrrrrr}
\hline
Catalogue name & Source	&	MJD	&	$\alpha_1$	&	$\alpha_2$	&	$\alpha_3$	&	$\alpha_4$ &	$\alpha_5$		\\
(1) & (2) & (3) & (4) & (5) & (6) & (7) & (8)\\
\hline
PKS 0440$-$00 & 0440$-$003	&	56543	&				&	$-0.45\pm0.01	$		&	$-0.54\pm0.01	$		&	$-0.63\pm0.03 $			&	$-0.51\pm0.02	$		\\
&	&	60321	&				&	$-0.14\pm 0.01	$		&	$-0.34\pm 0.02  $			&	$-0.41\pm 0.07    $			&	$-0.57\pm0.04	$		\\
PMN J0948+0022 & 0946+006	&	59240	&				&	0.10 $\pm$ 0.25			&	0.80 $\pm$ 0.23			&	0.69 $\pm$ 0.74			&	0.12 $\pm$ 0.31			\\
&	&	59525	&				&	1.38 $\pm$ 0.07			&	0.48 $\pm$ 0.05			&	0.27 $\pm$ 0.17			&				\\
PKS 1244$-$255 & 1244$-$255	&	56550	&				&	0.31 $\pm$ 0.02			&	0.34 $\pm$ 0.01			&	0.38 $\pm$ 0.04			&				\\
&	&	58420	&	0.97 $\pm$ 0.01			&	0.99 $\pm$ 0.01			&	0.40 $\pm$ 0.01			&	0.39 $\pm$ 0.04			&				\\
PKS 1502+036 & 1502+036	&	55045	&	0.79 $\pm$ 0.15			&	$-$0.07 $\pm$ 0.06			&	$-$0.15 $\pm$ 0.04			&	$-$0.32 $\pm$ 0.12			&				\\
&	&	60000	&				&	0.04 $\pm$ 0.03			&	$-$0.49 $\pm$ 0.04			&	$-$0.47 $\pm$ 0.17			&	$-$0.11 $\pm$ 0.09			\\
PKS 2004$-$447 & 2004$-$447	&	57349	&				&	$-$0.67 $\pm$ 0.04			&	$-$0.80 $\pm$ 0.05			&	$-$0.86 $\pm$ 0.23			&	$-$0.75 $\pm$ 0.18			\\
&	&	59989	&				&	0.09 $\pm$ 0.06			&	$-$0.16 $\pm$ 0.10			&	$-$0.32 $\pm$ 0.41			&				\\
\hline
\multicolumn{8}{l}{$Notes$. Column (1) source name in catalogues; column (2) is source names in project C007; column (3) are MJD time of epochs;}  \\ 
\multicolumn{8}{l}{ column (4-8) are the in-band spectral index between 2.1 and 5.5\,GHz, between 5.5 and 9.0\,GHz, between 9.0 and 17\,GHz, between} \\
\multicolumn{8}{l}{ 17 and 21\,GHz and between 21 and 33\,GHz, respectively.} \\
\end{tabular}
\end{table*}

\begin{table*}
\setlength\tabcolsep{4pt}
\caption{Flux densities and spectral indices of NLS1s.}
\rotatebox{90}{%
\label{tab:flux_spectral_index}
\begin{tabular}{lccccccccccccccl}
\hline
Name & RA & Dec &  $z$ & $F_{887.5\,\rm MHz}$   &   $F_{1367.5\,\rm MHz}$	& Ep & $T$ & $F_{5.5\,\rm GHz}$	&$F_{9.0\,\rm GHz}$ &  $S_{5.5\, \rm GHz}$ &$S_{9.0\, \rm GHz}$ & $\alpha_{1}$ & $\alpha_{2}$ &	  $\alpha_{3}$ & Note   \\
  &   (h:m:s)   & (d:m:s) &  &   (mJy)     & (mJy) &   &(min) &    (mJy)	& (mJy)	   &   (mJy)	& (mJy)    &      &     	\\
 (1)  & (2) & (3) & (4) & (5) & (6) & (7) & (8) & (9) & (10) & (11)  & (12) & (13) & (14) &(15) & (16)\\
\hline

J0122$-$2646	& 	01:22:37.52 & $-$26:46:45.9 &	0.417 	&	2.8 	$\pm$	0.9 	&	2.0 	$\pm$	0.3 	&	D	&	30	&	7.1 	$\pm$	0.2 	&	1.1 	$\pm$	0.3 	&	$	0.9 	\pm	0.1 	^v	$	&             -                 &$	-0.8 	\pm	0.8 	$	&	$	0.9 	\pm	0.1 	$	&	$	-3.8 	\pm	0.7 	$	&	SFS \\ 
J0133$-$2109	&	01:33:35.03	& $-$21:09:59.0 & 0.130 	&	3.5 	$\pm$	0.5 	&	2.3 	$\pm$	0.4 	&	B	&	20	&<	0.3 		&<	0.3 		&	-	&             -                 &$	-1.0 	\pm	0.5 	$	&<	$	-1.4 	 	$	&	-	&	SS~-	\\
J0230$-$0859	&02:30:05.53 & $-$08:59:53.20 &	0.016 	&	4.7 	$\pm$	0.6 	&	3.2 	$\pm$	0.4 	&	F	&	45	&	0.6 	$\pm$	0.1 	&<	0.3 	&	$	1.2 	\pm	0.0 	^v	$	&             -                 &$	-0.9 	\pm	0.4 	$	&	$	-1.2 	\pm	0.2 	$	&<	$	-1.3 		$	&	SSS	\\
J0307$-$7250	&03:07:35.32& $-$72:50:02.50& 	0.028 	&		-		&		-		&	B	&	30	&<	0.3 		&<	0.3 		&<	$	0.1 	^a	$	&$	0.4	\pm	0.03	^a	$&   -	&	-	&	-	&	-	\\
J0354$-$1340	&03:54:32.85& $-$13:40:07.29&	0.076 	&	18.1 	$\pm$	1.3 	&	6.6 	$\pm$	0.5 	&	D	&	30	&	7.0 	$\pm$	0.1 	&	8.9 	$\pm$	0.0 	&	$	5.1 \pm 0.0	 	^v	$	&             -                 &$	-2.3 	\pm	0.2 	$	&	$	0.0 	\pm	0.1 	$	&	$	0.5 	\pm	0.0 	$	&	SFF	\\
J0400$-$2500	&04:00:24.42& $-$25:00:44.48&	0.097 	&	4.0 	$\pm$	0.4 	&	2.3 	$\pm$	0.3 	&	D	&	30	&	0.7 	$\pm$	0.2 	&<	0.3 			&	$	1.2 	\pm	0.0 	^v	$	&             -                 &$	-1.2 	\pm	0.4 	$	&	$	-0.8 	\pm	0.2 	$	&<	$	-2.0 		$	&	SSS	\\
J0422$-$1854	&04:22:56.58& $-$18:54:41.39&	0.064 	&	4.3 	$\pm$	0.7 	&	1.7 	$\pm$	0.2 	&	F	&	45	&	0.7 	$\pm$	0.2 	&<	0.4 			&	$	1.1 	\pm	0.0 	^v	$	&             -                 &$	-2.1 	\pm	0.5 	$	&	$	-0.6 	\pm	0.2 	$	&<	$	-1.3 	 	$	&	SSS	\\
J0436$-$1022	&04:36:22.32& $-$10:22:33.23&	0.036 	&	21.2 	$\pm$	1.0 	&	21.3 	$\pm$	2.1 	&	D	&	30	&	4.1 	$\pm$	0.3 	&	2.1 	$\pm$	0.1 	&	$	4.6 	\pm	0.2 	^v	$	&             -                 &$	0.0 	\pm	0.2 	$	&	$	-1.2 	\pm	0.1 	$	&	$	-1.3 	\pm	0.2 	$	&	FSS	\\
J0447$-$0508	&04:47:20.72& $-$05:08:13.99&	0.045 	&	89.0 	$\pm$	6.8 	&	8.2 	$\pm$	0.6 	&	F	&	45	&	2.4 	$\pm$	0.2 	&	1.6 	$\pm$	0.3 	&	$	4.0 	\pm	0.1 	^v	$	&             -                 &$	-5.5 	\pm	0.3 	$	&	$	-0.9 	\pm	0.1 	$	&	$	-0.8 	\pm	0.4 	$	&	SSS	\\
J0452$-$2953	&04:52:30.13& $-$29:53:35.59&	0.247 	&	19.1 	$\pm$	1.2 	&	10.2 	$\pm$	0.7 	&	D	&	30	&	0.7 	$\pm$	0.1 	&	0.9 	$\pm$	0.1 	&	$	3.4 	\pm	0.0 	^v	$	&             -                 &$	-1.4 	\pm	0.2 	$	&	$	-1.9 	\pm	0.1 	$	&	$	0.6 	\pm	0.4 	$	&	SSF	\\
J0549$-$2425	&05:49:14.89& $-$24:25:51.50&	0.045 	&	5.1 	$\pm$	0.6 	&	3.7 	$\pm$	0.4 	&	F	&	45	&<	0.5 		&<	0.2 		&	$	1.6 	\pm	0.0 	^v	$	&             -                 &$	-0.7 	\pm	0.4 	$	&<	$	-1.4 	$	&	-	&	SS~-	\\
J0609$-$5606	&06:09:17.48& $-$56:06:58.40&	0.032 	&	4.1 	$\pm$	0.5 	&	3.1 	$\pm$	0.3 	&	B	&	30	&<	0.3 		&<	0.3 		&	$	0.4 	\pm	0.0 	^a	$	&<	$0.12			^a	$&$	-0.7 	\pm	0.4 	$	&<	$	-1.8 	$	&	-	&	SS~-	\\
J1044$-$1826	&10:44:48.70& $-$18:26:51.53&	0.113 	&	6.8 	$\pm$	0.7 	&	5.4 	$\pm$	0.6 	&	E	&	45	&	1.4 	$\pm$	0.2 	&	1.9 	$\pm$	0.1 	&	$	1.6 	\pm	0.0 	^v	$	&             -                 &$	-0.5 	\pm	0.4 	$	&	$	-1.0 	\pm	0.1 	$	&	$	0.7 	\pm	0.4 	$	&	SSF	\\
J1057$-$4039	&10:57:27.87& $-$40:39:40.60&	0.398 	&	475.4 	$\pm$	2.5 	&	309.2 	$\pm$	18.6 	&	E	&	30	&	282.6 	$\pm$	0.1 	&	301.2 	$\pm$	0.1 	&	$	167.7 	\pm	0.1 	^a	$	&	$151.7	\pm	0.07	^a	$&$	-1.0 	\pm	0.1 	$	&	$	-0.1 	\pm	0.0 	$	&	$	0.1 	\pm	0.0 	$	&	SFF	\\
J1500$-$7248	&15:00:12.81& $-$72:48:40.30&	0.141 	&	16.8 	$\pm$	0.4 	&	13.4 	$\pm$	0.9 	&	E	&	30	&	54.1 	$\pm$	0.1 	&	60.1 	$\pm$	0.1 	&	$	31.4 	\pm	0.0 	^a	$	&	$35.4	\pm	0.04	^a	$&$	-0.5 	\pm	0.2 	$	&	$	1.0 	\pm	0.0 	$	&	$	0.2 	\pm	0.0 	$	&	SFF	\\
J1500$-$7248	&15:00:12.81& $-$72:48:40.30&	0.141 	&	16.8 	$\pm$	0.4 	&	13.4 	$\pm$	0.9 	&	A	&	5	&	78.7 	$\pm$	0.1 	&	76.9 	$\pm$	0.1  	&	$	31.4 	\pm	0.0 	^a	$	&	$35.4	\pm	0.04	^a	$&$	-0.5 	\pm	0.2 	$	&	$	1.3 	\pm	0.0 	$	&	$	0.0 	\pm	0.0   	$	&	SFF	\\
J1515$-$7820	&15:15:15.20& $-$78:20:12.00&	0.259 	&	25.1 	$\pm$	0.5 	&	16.3 	$\pm$	1.0 	&	E	&	30	&	1.8 	$\pm$	0.6 	&	1.8 	$\pm$	0.1 	&	$	3.9 	\pm	0.0 	^a	$	&	$2.1	\pm	0.03	^a	$&$	-1.0 	\pm	0.2 	$	&	$	-1.6 	\pm	0.2 	$	&	$	0.0 	\pm	0.7 	$	&	SSF	\\
J1515$-$7820	&15:15:15.20& $-$78:20:12.00&	0.259 	&	25.1 	$\pm$	0.5 	&	16.3 	$\pm$	1.0 	&	A	&	5	&	1.2 	$\pm$	0.3 	&	0.7 	$\pm$	0.2 	&	$	3.9 	\pm	0.0 	^a	$	&	$2.1	\pm	0.03	^a	$&$	-1.0 	\pm	0.2 	$	&	$	-1.9 	\pm	0.2 	$	&	$	-1.1 	\pm	0.7 	$	&	SSS	\\

J2137$-$1112	&21:37:47.95& $-$11:12:03.60& 	0.113 	&		-		&	0.9 	$\pm$	0.3 	&	B	&	20	&<	0.3 	&<	0.2 	&	-	&             -                 &-	&	<$	-0.7 	$	&	-	&	-~S~-	\\
J2229$-$1401	&22:29:03.51& $-$14:01:06.20&	0.236 	&		-		&			-	&	C	&	20	&	2.2 	$\pm$	0.1 	&<	0.4 		&	-	&             -                 &-	&	-	&<	$	-3.6 		$	&	-~-~S	\\
J2244$-$1822	&22:44:58.19& $-$18:22:49.50&	0.198 	&		-		&			-	&	C	&	20	&<	0.6 		&<	0.4 		&	-	&             -                 &-	&	-	&	-	&	-	\\
J2250$-$1152	&22:50:14.06& $-$11:52:00.80&	0.118 	&		-		&			-	&	C	&	20	&	4.2 	$\pm$	0.1 	&	1.0 	$\pm$	0.4 	&	-	&             -                 &-	&	-	&	$	-3.0 	\pm	0.8 	$	&	-~-~S	\\
J2311$-$2022	&23:11:03.26& $-$20:22:20.60&	0.121 	&		-		&			-	&	C	&	30	&<	0.3 		&<	0.4 		&	-	&             -                 &-	&	-	&	-	&	-	\\
																
Median	&-	& - & 0.113 	&	6.8 	$\pm$	0.7 	&	4.6 	$\pm$	0.4 	&	-	&	30	&	2.3 	$\pm$	0.1 	&	1.9 	$\pm$	0.1 	&	$	3.4 	\pm	0.0 		$	&$	18.8	\pm	0.0		$&$	-1.0 	\pm	0.5 	$	&	$	-0.9 	\pm	0.0 	$	&	$	0.0 	\pm	0.3  $	&	-	\\

\hline
\multicolumn{16}{l}{$Notes$. Column (1): source names; column (2-3): the coodinate of sources;column (4): redshifts from NED (\url{http://ned.ipac.caltech.edu/}); column (5-6): flux densities at 887.5 MHz (RACS-Low); }\\%
\multicolumn{16}{l}{and at 1367.5 MHz (RACS-Mid), respectively; column (7): code of epochs, letter A, B, C, D, E and F represent the observations on Dec 3, 2023 and Jan 5, 2024 with 6D Array, on Jan 12 and Jan 25, 2024}\\
\multicolumn{16}{l}{with EW367 Array, and Feb 17 and Feb 28, 2024 with 6A Array, respectively; column (8): exposure time; column(9-10): flux densities at 5.5 GHz (ATCA) and 9.0 GHz (ATCA), column (11-12): previous}\\
\multicolumn{16}{l}{flux densities: superscript letter $^a$ represents from ATCA observations in 4cm band \citep{2022MNRAS.512..471C} and superscript letter $^v$ represents from VLA observations at 5.5 GHz \citep{2020MNRAS.498.1278C}; Columns }\\
\multicolumn{16}{l}{(13-15): in-band spectral indices between 887.5 MHz and 1367.5 MHz, between 1367.5 MHz and 5.5 GHz (new data), and between 5.5\,GHz (new data) and 9.0\,GHz (new data), respectively; column(16): }\\
\multicolumn{16}{l}{spectral types of $\alpha_1$, $\alpha_2$ and $\alpha_3$, respectively. The last line gives the median of flux densities and in-band spectral indices at different bands, for sources J1500$-$7248 and J1515$-$7820 we select the longer }\\
\multicolumn{16}{l}{integrated time (30 mins) for the median calculation.}
\end{tabular}
}
\end{table*}

\section*{Acknowledgements}

We give great thanks to the referee for comments which have  improved the paper.
XS is grateful for the support from CSIRO and the SKA PhD scholarship from China Scholarship Council.
The Australia Telescope Compact Array is part of the Australia Telescope National Facility (https://ror.org/05qajvd42) which is funded by the Australian Government for operation as a National Facility managed by CSIRO. We acknowledge the Gomeroi people as the Traditional Owners of the Observatory site.

XS and MFG are supported by the National Science Foundation of China (grant 12473019), the National SKA Program of China (Grant
No. 2022SKA0120102), the Shanghai Pilot Program for Basic Research-Chinese
Academy of Science, Shanghai Branch (JCYJ-SHFY-2021-013), and the China Manned Space
Project with No. CMSCSST-2021-A06.

\section*{Data Availability}

The ATCA data used in this paper is available in the Australia Telescope Online Archive (ATOA) at \url{ht tps://at oa.atnf.csiro.au} with project codes of CX540, C007 and C1730.


\newpage
\bibliographystyle{mnras}
\bibliography{example} 


\bsp	
\label{lastpage}
\end{document}